\documentclass[english]{article}
\usepackage[T1]{fontenc}
\usepackage[latin9]{inputenc}
\usepackage{geometry}
\usepackage{babel}
\usepackage{float}
\usepackage{amsmath}
\usepackage{amssymb}
\usepackage[unicode=true,pdfusetitle,
 bookmarks=true,bookmarksnumbered=false,bookmarksopen=false,
 breaklinks=false,pdfborder={0 0 0},pdfborderstyle={},backref=false,colorlinks=false]
 {hyperref}

\makeatletter

\floatstyle{ruled}
\newfloat{algorithm}{tbp}{loa}
\providecommand{\algorithmname}{Algorithm}
\floatname{algorithm}{\protect\algorithmname}

\usepackage{graphicx}

\@ifundefined{showcaptionsetup}{}{%
 \PassOptionsToPackage{caption=false}{subfig}}
\usepackage{subfig}
\makeatother

\usepackage{listings}

\begin{document}
\title{Meshless discretization of the discrete-ordinates transport equation
with integration based on Voronoi cells\vspace{0.1\textheight}
}
\author{Brody R. Bassett$^{a,\star}$, J. Michael Owen$^{a}$\\
\\
{\normalsize{}\href{mailto:bassett4@llnl.gov}{bassett4@llnl.gov},
\href{mailto:mikeowen@llnl.gov}{mikeowen@llnl.gov}}\\
{\normalsize{}}\\
{\normalsize{}$^{a}$Lawrence Livermore National Laboratory}\\
{\normalsize{}7000 East Avenue, Livermore, CA, 94550}\\
{\normalsize{}}\\
{\normalsize{}$^{\star}$Corresponding author}\vspace{0.1\textheight}
}
\date{$ $}
\maketitle
\begin{abstract}
The time-dependent radiation transport equation is discretized using
the meshless-local Petrov-Galerkin method with reproducing kernels.
The integration is performed using a Voronoi tessellation, which creates
a partition of unity that only depends on the position and extent
of the kernels. The resolution of the integration automatically follows
the particles and requires no manual adjustment. The discretization
includes streamline-upwind Petrov-Galerkin stabilization to prevent
oscillations and improve numerical conditioning. The angular quadrature
is selectively refineable to increase angular resolution in chosen
directions. The time discretization is done using backward Euler.
The transport solve for each direction and the solve for the scattering
source are both done using Krylov iterative methods. Results indicate
first-order convergence in time and second-order convergence in space
for linear reproducing kernels. \\
\end{abstract}


\pagebreak{}

\section{Introduction\label{sec:Introduction}}

Meshless methods have a rich variety of applications in hydrodynamic
modeling, particularly in situations that are challenging to mesh
initially or where complex flows make it difficult to maintain a good
meshed description. Examples include problems with large deformations,
fractures, unstable flows, and a variety of astrophysical problems
(where such methods originated) \cite{belytschko1996meshless}. However,
many astrophysical problems in the high energy density physics regime
also require a thermal radiation transfer treatment \cite{castor2004radiation},
and to date there has been much less work on meshfree treatments for
radiation transport. One fundamental choice one must make is what
sort of angular representation of the radiation is appropriate for
the physics problem at hand. The few prior meshfree thermal radiative
transfer treatments have generally focused on radiation diffusion
and similar approximations \cite{whitehouse2004smoothed,whitehouse2005faster,viau2006implicit,mayer2007fragmentation,petkova2009implementation},
wherein the angular distribution of the radiation is neglected. This
is appropriate for situations dominated by scattering and absorption
(such as deep in stellar interiors), but many interesting problems
include both transparent and opaque regions. To model these problems
accurately, a more complicated angular discretization of the radiation
transport equation is needed. The goal of this work is to develop
a discrete ordinates radiation transport implementation that is compatible
with a method such as smoothed particle hydrodynamics \cite{monaghan2005smoothed}
and variations of the reproducing kernel particle method \cite{liu1995reproducing,frontiere2017crksph}.
For this paper, the transport does not include radiation hydrodynamics
effects and the nonlinear emission terms in thermal radiative transfer. 

The discrete ordinates radiation transport equation has been solved
previously using meshless methods, including for collocation methods
\cite{sadat2006use,sadat2012meshless,kindelan2010application,liu2007least,zhao2013second,kashi2017mesh}
and the meshless local Petrov-Galerkin (MLPG) method \cite{liu2007meshless,bassett2019meshless}.
The collocation methods often use the second-order form of the radiation
transport equation, such as the self-adjoint angular flux equation
\cite{morel1999self}, while the MLPG methods use a background mesh
for the integration. 

This paper extends the past MLPG discretizations of the radiation
transport equation in several ways. First, reproducing kernels \cite{liu1995reproducing}
are used instead of moving least squares, which reduces the number
of linear solves needed to evaluate the kernels. Using these RK functions,
higher-than-second-order convergence is demonstrated for certain choices
of RK correction order. Second, the implementation is time-dependent,
which adds additional complexity and means that the integration, which
for for some problems may be performed at each time step, cannot require
manual adjustment and needs to be efficient. Third, the integration
is done using a Voronoi decomposition, which meets both of these criteria.
Fourth, the angular quadrature can be refined, which permits consideration
of problems that require high angular resolution in specific directions.
Finally, the implementation of the code has been done inside a code
that already includes radiation hydrodynamics with diffusion \cite{bassett2020efficient,bassett2020efficienta}
and reproducing kernel hydrodynamics \cite{frontiere2017crksph},
which should permit future consideration of meshless radiation hydrodynamics. 

In many cases, including the meshless Galerkin approach \cite{li2002meshfree},
the integration is done using a background mesh \cite{nguyen2008meshless}.
In the original MLPG paper, the authors recommend using circular (2D)
or spherical (3D) domains of integration to retain a truly meshless
method \cite{atluri1998new}. Integration can also be performed by
introducing a quadrature into the lens-shaped intersection of two
kernels \cite{racz2012novel} or by reducing the dimensionality of
the integrals \cite{khosravifard2010new}. One method of avoiding
any evaluations outside of the kernel centers is by using nodal integration,
which may require stabilization for the derivatives \cite{chen2001stabilized}
and cannot be further refined to increase accuracy \cite{zhou2003nodal},
although using a Voronoi tessellation to create the volumes can increase
accuracy \cite{puso2008meshfree}. By using a quadrature within a
Voronoi tessellation to integrate the kernels, the resolution of the
integrals follows the resolution of the particles and that the integration
mesh depends only on the location of the particles and the boundary
surfaces of the problem. This, in effect, makes the integration invariant
to the rotation or translation of the points within the domain. 

The paper is structured as follows. In Sec. \ref{sec:Interpolation},
the smoothed particle hydrodynamics (SPH) and reproducing kernels
(RK) are introduced, along with methods for calculating derivatives.
Next, the integration method using a Voronoi tessellation is discussed
in Sec. \ref{sec:Integration} and transformations from local to global
coordinates are derived in App. \ref{sec:Transformations}. The time-dependent
transport equation is introduced and discretized using these kernels
and integration methods in Sec. \ref{sec:Transport}. The discretization
is then tested in Sec. \ref{sec:Results} for a purely absorbing problem,
two manufactured solutions, a purely-scattering problem with disparate
cross sections, and a problem with a source in a void far from a strong
absorber whose solution is derived in App. \ref{sec:Analytic-asteroid}.
Finally, conclusions and future work are discussed in Sec. \ref{sec:Conclusions}. 

\section{Interpolation methods\label{sec:Interpolation}}

This section introduces the smoothed particle hydrodynamics (SPH)
and reproducing kernel (RK) functions. Here and throughout this paper,
Greek superscripts represent dimensional indices, with repeated indices
representing summation, e.g. $x^{\alpha}$ is the component $\alpha$
of the vector $x$ and $X^{\alpha\beta}$ is the component $\alpha,\beta$
of the tensor $X$. The notation $\partial_{x}^{\alpha}$ denotes
the partial derivative with respect to $x^{\alpha}$. Subscripts denote
evaluation of a function at a discrete point, e.g. $f_{j}=f\left(x_{j}\right)$,
unless otherwise noted. 

\subsection{Introduction to smoothed particle hydrodynamics\label{subsec:SPH}}

In SPH, spatial fields are interpolated from the discrete values at
individual points using functions referred to as kernels, which are
symmetric functions (typically of vector distance) with compact support,
i.e., they fall to zero at some range. The support of the kernel is
determined by a smoothing parameter, which translates from physical
space to the reference space for the kernel. This smoothing parameter
can either be a scalar $h$, as in standard SPH, or a symmetric tensor
$H^{\alpha\beta}$, as in adaptive smoothed particle hydrodynamics
(ASPH) \cite{owen1998adaptive}. This smoothing parameter is generally
allowed to vary point to point, so $H_{i}^{\alpha\beta}$in general.
Note that when using SPH, the interpolation kernel is radially symmetric,
while under ASPH this is not necessarily true. Because the ASPH $H^{\alpha\beta}$
tensor is symmetric, the corresponding smoothing scale isocontours
around a point are elliptical (2D) or ellipsoidal (3D). Using the
tensor smoothing parameter (which has units of inverse length), the
transformed distance vector in ASPH reference space $\eta\left(x\right)$
and the scaled distance $\chi\left(\eta\right)$ are 
\begin{gather}
\eta^{\alpha}=H^{\alpha\beta}x^{\beta},\label{eq:ASPH-eta}\\
\chi=\sqrt{\eta^{\alpha}\eta^{\alpha}}.
\end{gather}
Note that SPH is a simply a special case of ASPH, wherein $H^{\alpha\beta}=h^{-1}\delta^{\alpha\beta}$,
so in the SPH case Eq. (\ref{eq:ASPH-eta}) reduces to
\begin{equation}
\eta^{\alpha}=x^{\alpha}/h.
\end{equation}

With these conventions in mind the kernel equations can be written
in terms of $H^{\alpha\beta}$ and apply equally to SPH or ASPH. The
kernel $W\left(x\right)$ and its derivatives can be defined in terms
of the base kernel in reference space $W^{b}\left(\chi\right)$ as
\begin{gather}
W=W^{b},\\
\partial_{x}^{\gamma}W=\partial_{\eta}^{\alpha}\chi\partial_{x}^{\gamma}\eta^{\alpha}\partial_{\chi}W^{b}.
\end{gather}
The derivative equation can be simplified by inserting the derivatives
of $\chi$ and $\eta$,
\begin{gather}
\partial_{\eta}^{\alpha}\chi=\frac{\eta^{\alpha}}{\chi},\\
\partial_{x}^{\gamma}\eta^{\alpha}=H^{\alpha\gamma},
\end{gather}
which results in
\begin{gather}
\partial_{x}^{\gamma}W=\frac{\eta^{\alpha}}{\chi}H^{\alpha\gamma}\partial_{\chi}W^{b}.
\end{gather}
As the kernels approximate delta functions ($W\left(x\right)\to\delta\left(x\right)$
as $H^{\alpha\beta}\to\infty$), they can be used in interpolation,
\begin{align}
f\left(x\right) & =\int_{V}\partial\left(x-x'\right)f\left(x'\right)dV'\nonumber \\
 & \approx\int_{V}W\left(x-x'\right)f\left(x'\right)dV'.\label{eq:sph-delta-property}
\end{align}
The kernels are normalized such that formally the volume integral
is unity, 
\begin{equation}
\int_{V}W\left(x-x'\right)dV'=1,
\end{equation}
though this property is only approximately true in the discrete case
for SPH. The interpolant can be discretized for a set of these kernels
with discrete positions $x_{j}$ with associated volumes $V_{j}$,
\begin{equation}
\overline{f}\left(x\right)=\sum_{j}V_{j}W_{j}\left(x\right)f_{j},
\end{equation}
where 
\begin{equation}
W_{i}\left(x\right)=W\left(x-x_{i}\right)
\end{equation}
and $\overline{f}$ denotes the discrete interpolant of $f$. 

\subsection{Introduction to reproducing kernels\label{subsec:Reproducing}}

In general, the standard SPH kernels cannot reproduce even a constant
solution exactly,
\begin{equation}
\sum_{j}V_{j}W_{j}\left(x\right)\neq\text{const}.
\end{equation}
The SPH kernels can be augmented with RK functions \cite{liu1995reproducing},
which permit exact interpolation of functions up to a certain polynomial
order. Interpolation with RK functions $U_{i}\left(x\right)$ works
the same as in SPH,
\begin{equation}
\overline{f}\left(x\right)=\sum_{j}V_{j}U_{j}f_{j},\label{eq:rk-interpolant}
\end{equation}
with the caveat that the RK functions have the property that
\begin{equation}
\sum_{j}V_{j}U_{j}\left(x\right)f_{j}=f\left(x\right),\quad f\left(x\right)\in\mathbb{P}_{n}
\end{equation}
where $\mathbb{P}_{n}$ is the space of polynomials with degree less
than or equal to $n$. These functions and their derivatives are defined
in terms of the SPH functions as 
\begin{gather}
U_{i}=P_{i}^{\top}CW_{i},\\
\partial^{\gamma}U_{i}=\left(\partial_{x}^{\gamma}P_{i}^{\top}C+P_{i}\partial_{x}^{\gamma}C\right)W_{i}+P_{i}^{\top}C\partial_{x}^{\gamma}W_{i},
\end{gather}
where $P\left(x\right)$ is the polynomial basis vector,

\begin{gather}
P\left(x\right)=\left[1,x^{\alpha},x^{\alpha}x^{\beta},\cdots\right]^{\top},\\
P_{i}=P\left(x-x_{i}\right),\\
P_{ij}=P\left(x_{j}-x_{i}\right),
\end{gather}
and $C\left(x\right)=\left[C^{0}\left(x\right),C^{1}\left(x\right),C^{2}\left(x\right),\cdots\right]^{\top}$
is a corrections vector of the same size as the polynomial vector
with coefficients $C^{k}$ (i.e. the component $k$ of $C$) to be
determined. Suppose that $F$ is a vector of arbitrary coefficients.
The RK method calculates $C$ such that the reproducing kernels can
exactly represent $F^{\top}P$. The term $F^{\top}P_{i}$ is interpolated
as
\begin{align}
F^{\top}P_{i} & =\sum_{j}V_{j}F^{\top}P_{ji}U_{j}\nonumber \\
 & =F^{\top}\sum_{j}V_{j}P_{ji}P_{j}^{\top}CW_{j}.
\end{align}
This equation must be true for each component of $F$, 
\begin{equation}
P_{i}=\sum_{j}V_{j}P_{ji}P_{j}^{\top}CW_{j},
\end{equation}
and can be evaluated at the point $x_{i}$ to produce simple conditions
for the coefficients, 
\begin{equation}
\sum_{j}V_{j}P_{ji}P_{ji}^{\top}W_{ji}C_{i}=G,
\end{equation}
where $W_{ji}=W_{j}\left(x_{i}\right)$ and $G=\left[1,0,0,\cdots\right]^{\top}$.
The matrix for this linear system and its derivatives can be written
explicitly as 
\begin{gather}
M_{i}=\sum_{j}V_{j}P_{ji}P_{ji}^{\top}W_{ji},\label{eq:rk-matrix}\\
\partial^{\gamma}M_{i}=\sum_{j}V_{j}\left[\left(\partial^{\gamma}P_{ji}P_{ji}^{\top}+P_{ji}\partial^{\gamma}P_{ji}^{\top}\right)W_{ji}+P_{ji}P_{ji}^{\top}\partial^{\gamma}W_{ji}\right].
\end{gather}
In terms of these matrices, the linear systems to solve for $C_{i}$
and its derivatives can be written as
\begin{gather}
M_{i}C_{i}=G,\\
\partial^{\gamma}M_{i}C_{i}+M_{i}\partial^{\gamma}C_{i}=0.
\end{gather}
By first solving for $C_{i}$ and then $\partial^{\gamma}C_{i}$,
the only matrix that needs to be inverted is $M_{i}$, 
\begin{gather}
C_{i}=M_{i}^{-1}G,\\
\partial^{\gamma}C_{i}=-M_{i}^{-1}\partial^{\gamma}M_{i}C_{i},
\end{gather}
which lets us reuse its factorization. 

\section{Meshless integration\label{sec:Integration}}

In this section, the methodology for creating a Voronoi tessellation
is introduced, the meshless integration process is described, and
the connectivity for the weak-form kernels is derived in terms of
a similar strong-form connectivity. 

\subsection{Process of creating the Voronoi tessellation}

As discussed in the introduction (Sec. \ref{sec:Introduction}), there
have been several methods developed to integrate radial basis functions.
For these results, the problem is decomposed using what is essentially
a Voronoi tessellation constructed using the PolyClipper library \cite{owen2020polyclipper},
with one line segment (1D), polygon (2D), or polyhedron (3D) per meshfree
point. Each cell is then further decomposed into triangles (2D) or
tetrahedra (3D), with surfaces defined by points (1D), line segments
(2D) or triangles (3D). It is worth pointing out that the decomposition
is not truly the Voronoi. Rather the decomposition begins with an
initial polytope for each point that encompasses the finite kernel
extent of that point (i.e., the space over which its kernel value
is non-zero), which is progressively clipped by planes halfway between
the point in question and each neighbor point it interacts with. In
the end, this results in a tiling of space with these polytopes per
point that exactly constructs a partition of unity in space for all
points that overlap. 

Note that the topological connection between the polytopes for each
point is not computed, but only a unique polygon or polyhedron for
each point independently. Figure \ref{fig:polygon-volume}a shows
a cartoon of this process. The goal is to construct the Voronoi-like
polygon for the central red point, which has a set of neighbor points
it overlaps (in blue), and a non-zero kernel extent represented as
the gray region. The starting polygon for this point is the bounding
surface of this gray region, which is progressively clipped by planes
half-way between the central red point and each of its neighbors.
In the end all that is left is the central light red polygon, which
is the unique volume closer to the red point than any of its neighbors.
To facilitate simple integration quadratures, these polytopes for
each point are further broken down into triangles (in 2D) and tetrahedra
(in 3D). Fig. \ref{fig:polygon-decomposition} shows this procedure
for the polygon generated in Fig. \ref{fig:polygon-volume}, where
the cross markers denote the centroid of each of the sub-triangles
in the polygon. Note the centroid of the polygon does not necessarily
coincide with the original point used to construct it.

\subsection{Meshless integration quadrature\label{subsec:Meshless-quadrature}}

The set of integration quadratures over the base shapes represents
a contiguous, non-overlapping quadrature that covers the domain. A
Gauss-Legendre quadrature is used for integration of the line segments,
while symmetric quadrature rules as described in Ref. \cite{witherden2015identification}
are used to integrate the triangles and tetrahedra. Appendix \ref{sec:Transformations}
presents information on how the integrals are transformed from physical
to reference space. 

At each integration point, all the functions whose support includes
the integration point must be evaluated. The RK functions (Sec. \ref{subsec:Reproducing})
are expensive to evaluate relative to a standard SPH kernel and the
evaluation of one such function depends on the values of all other
functions at that point. As such, a large amount of computation can
be saved by making the quadrature the outermost loop in the code,
as shown in Alg. \ref{alg:integration}. For each quadrature point,
all functions whose support includes the integration quadrature point
are evaluated. Then these values are used to perform each integral. 

It is worth comparing this approach with prior background integration
methodologies, wherein a traditional background mesh (often some sort
of orthogonal Cartesian grid aligned with the lab frame) is placed
independently of the meshless points, even if the point locations
inform characteristics of the background mesh. The approach in this
section, using a Voronoi tessellation for the integration, produces
an integration mesh whose properties are solely determined by the
volume and relative positioning of the meshless points, which makes
it invariant to rotation or translation. This integration does not
meet the strictest of meshless criteria, in which no mesh is allowed
\cite{atluri1999critical}, but does meet a looser criterion in that
all information can be derived directly from the meshless points.
The geometry of each point's unique volume is constructed based solely
on the positions of surrounding points without storing, evolving,
or specifying anything except the the point positions and kernel extents.

\subsection{Strong and weak forms for reproducing kernels}

For the following sections, to simplify the notation, volume and surface
integrals will be written as 
\begin{gather}
\left\langle f,g\right\rangle =\int_{V}fgdV,\\
\left(f,g\right)=\int_{S}fgdS.
\end{gather}
For many applications, such as hydrodynamics and diffusion, SPH and
RK can be used to directly discretize the equations via collocation
\cite{monaghan2005smoothed}. The equation, in this example
\begin{equation}
a^{\alpha}\partial_{x}^{\alpha}f+bf=0,
\end{equation}
is first integrated by parts,
\begin{equation}
\left(U_{i},n^{\alpha}a^{\alpha}f\right)-\left\langle \partial_{x}^{\alpha}U_{i},a^{\alpha}f\right\rangle +\left\langle U_{i},bf\right\rangle =0,
\end{equation}
(with $n$ denoting the surface normal), the surface term is discarded,
and interpolants {[}Eq. (\ref{eq:rk-interpolant}){]} and the delta
function property {[}Eq. (\ref{eq:sph-delta-property}){]} are used
to simplify the equation to 
\begin{equation}
-\sum_{j}V_{j}a_{j}^{\alpha}f_{j}\partial_{x_{i}}^{\alpha}U_{ji}+b_{i}f_{i}=0.
\end{equation}
While this form of the equation has the advantage of simplicity, it
depends on a one-point quadrature rule for the integration,
\begin{equation}
\int_{V}U_{i}f\left(x\right)\approx f_{i},
\end{equation}
and generally throws away surface terms. Another option, and the one
used in this paper, is to insert a basis function expansion, 
\begin{equation}
f\left(x\right)=\sum_{j}V_{j}U_{j}g_{j}\label{eq:basis-expansion}
\end{equation}
(with coefficients $g_{j}$),
\begin{equation}
\sum_{j}V_{j}\left[\left(U_{i},n^{\alpha}a^{\alpha}U_{j}\right)-\left\langle \partial_{x}^{\alpha}U_{i},a^{\alpha}U_{j}\right\rangle +\left\langle U_{i},bU_{j}\right\rangle \right]g_{j}=0,
\end{equation}
 and perform the integrals directly using a quadrature like the one
described in Sec. \ref{subsec:Meshless-quadrature}, 
\begin{equation}
\left\langle \partial_{x}^{\alpha}U_{i},a^{\alpha}U_{j}\right\rangle \approx\sum_{k}w_{k}\partial_{x_{k}}^{\alpha}U_{ik}a_{k}^{\alpha}U_{jk},
\end{equation}
where $w_{k}$ are the weights of a quadrature spanning the integration
volume. For functions with compact support, this is the Meshless-Local
Petrov-Galerkin (MLPG) method. 

\subsection{Meshless connectivity}

Two types of connectivity are used in evaluating and storing the MLPG
integrals. In a standard SPH code, the connectivity is the sets of
points whose evaluation is nonzero at the center of the other, so
points $i$ and $j$ are neighbors if the support of $W_{i}$ includes
the point $x_{j}$ or vice versa, or 
\begin{equation}
\chi\left(\eta\left(x_{i}-x_{j}\right)\right)\leq r,\label{eq:sph-connectivity}
\end{equation}
where $r$ is the dimensionless support radius of the kernel. For
MLPG, the connectivity (or sets of points for which the bilinear integrals
are nonzero) is determined by whether points have overlapping support,
so points $i$ and $j$ are neighbors if the support regions of $W_{i}$
and $W_{j}$ intersect. The overlap connectivity is a subset of the
set of points for which 
\begin{equation}
\chi\left(\eta\left(x_{i}-x_{j}\right)\right)\leq2r.\label{eq:mlpg-connectivity}
\end{equation}
To show this, suppose that there is a point $x_{k}$ that is in the
support radius of $W_{i}$ and $W_{j}$. It follows from Eq. (\ref{eq:sph-connectivity})
and the triangle inequality in Euclidean space that $\chi\left(\eta\left(x_{i}-x_{j}\right)\right)\leq\chi\left(\eta\left(x_{i}-x_{k}\right)\right)+\chi\left(\eta\left(x_{k}-x_{j}\right)\right)\leq2r$. 

Standard SPH connectivity information can be used to both create the
overlap connectivity and calculate which functions are nonzero at
each integration point, which is similar to standard SPH connectivity.
If the radius of the MLPG kernels is doubled (or the smoothing length
altered to produce a similar effect), then by Eqs. (\ref{eq:sph-connectivity})
and (\ref{eq:mlpg-connectivity}), the standard SPH connectivity of
the doubled-radius will include all overlap neighbors. As each cell
produced by the Voronoi tessellation is completely contained within
the support of its associated MLPG point, this same double-radius
SPH connectivity for $i$ and $j$ will include all integration points
in the cell $i$ for which the kernel $W_{j}$ is nonzero. This is
why the same connectivity can be used for both the overlap of two
kernels and the overlap of a kernel with a Voronoi cell associated
with a kernel in Alg. \ref{alg:integration}. 

\section{Radiation transport\label{sec:Transport}}

In this section, the radiation transport equation is introduced and
discretized using MLPG with RK functions. Then, the iterative methods
that are used for the solution of the discretized equation are described.
Finally, the angular quadrature with selective refinement is introduced. 

\subsection{Discretization of the transport equation}

The gray radiation transport equation, which is the transport equation
integrated over all energies, is 
\begin{equation}
\partial_{t}\psi+\Omega^{\alpha}\partial_{x}^{\alpha}\psi+\sigma_{t}\psi=\frac{1}{4\pi}\sigma_{s}\phi+q,\label{eq:transport-equation}
\end{equation}
with the boundary condition 
\begin{equation}
\psi=\psi^{b},\quad x\in\partial V,\ \Omega^{\alpha}n^{\alpha}>0,\label{eq:rad-boundary}
\end{equation}
and initial condition 
\[
\psi=\psi^{\text{init}},\quad t=0,
\]
where $\Omega$ is the radiation propagation direction, $\psi$ is
the angular flux, $\phi=\int_{4\pi}\psi d\Omega$ is the scalar flux,
$\sigma_{s}$ is the scattering cross section, $\sigma_{a}$ is the
absorption cross section, $\sigma_{t}=\sigma_{a}+\sigma_{s}$ is the
total cross section, $q$ is a source that may include physics such
as thermal emission, $\psi^{b}$ is the incoming flux at the boundary,
$\psi^{\text{init}}$ is the initial angular flux, and $\partial V$
denotes the boundary of the domain. The discrete-ordinates approximation
evaluates this equation at discrete angles that are ordinates of a
quadrature over a unit sphere. Denoting these ordinates as $\Omega_{m}$
for the angular index $m$, integrals over the unit sphere become
\begin{equation}
\int_{4\pi}fd\Omega\approx\sum_{m}w_{m}f_{m},
\end{equation}
where $w_{m}$ are the weights of the quadrature. The transport equation
evaluated at the discrete ordinate $m$ becomes 
\begin{equation}
\partial_{t}\psi_{m}+\Omega_{m}^{\alpha}\partial_{x}^{\alpha}\psi_{m}+\sigma_{t}\psi_{m}=\frac{1}{4\pi}\sigma_{s}\phi+q_{m},
\end{equation}
with the scalar flux $\phi=\sum_{m}w_{m}\psi_{m}$. The backward Euler
(or fully-implicit) method is used to discretize in time, 
\begin{equation}
\frac{1}{c\Delta t}\psi_{m}+\Omega_{m}^{\alpha}\partial_{x}^{\alpha}\psi_{m}+\sigma_{t}\psi_{m}=\frac{1}{c\Delta t}\psi_{m}^{n-1}+\frac{1}{4\pi}\sigma_{s}\phi+q_{m},
\end{equation}
where all variables are evaluated at time index $n$ except where
noted otherwise as a superscript. 

A standard Galerkin approach to transport would be to multiply the
transport equation by $U_{i}$ and then integrate over the support
of $U_{i}$. For streamline-upwind Petrov-Galerkin (SUPG) stabilization,
the transport equation is instead multiplied by $U_{i}+\tau_{i}\Omega^{\alpha}\partial_{x}^{\alpha}U_{i}$,
where $\tau$ is a proportionality constant with unit length that
is chosen to be constant for each trial function. Performing this
operation, the transport equation becomes 
\begin{align}
 & \frac{1}{c\Delta t}\left\langle U_{i},\psi_{m}\right\rangle +\frac{1}{c\Delta t}\tau_{i}\Omega_{m}^{\alpha}\left\langle \partial_{x}^{\alpha}U_{i},\psi_{m}\right\rangle +\left(U_{i},\Omega_{m}^{\alpha}n^{\alpha}\psi_{m}\right)_{\Omega^{\alpha}n^{\alpha}>0}-\Omega_{m}^{\alpha}\left\langle \partial_{x}^{\alpha}U_{i},\psi_{m}\right\rangle \nonumber \\
 & \quad+\tau_{i}\Omega_{m}^{\alpha}\Omega_{m}^{\beta}\left\langle \partial_{x}^{\alpha}U_{i},\partial_{x}^{\beta}\psi_{m}\right\rangle +\left\langle U_{i},\sigma_{t}\psi_{m}\right\rangle +\tau_{i}\Omega_{m}^{\alpha}\left\langle \partial_{x}^{\alpha}U_{i},\sigma_{t}\psi_{m}\right\rangle \nonumber \\
 & \qquad=\frac{1}{c\Delta t}\left\langle U_{i},\psi_{m}^{n-1}\right\rangle +\frac{1}{c\Delta t}\tau_{i}\Omega_{m}^{\alpha}\left\langle \partial_{x}^{\alpha}U_{i},\psi_{m}^{n-1}\right\rangle +\left(U_{i},\left|\Omega_{m}^{\alpha}n^{\alpha}\right|\psi_{m}^{b}\right)_{\Omega^{\alpha}n^{\alpha}<0}\nonumber \\
 & \qquad\quad+\frac{1}{4\pi}\left\langle U_{i},\sigma_{s}\phi\right\rangle +\frac{1}{4\pi}\tau_{i}\Omega_{m}^{\alpha}\left\langle \partial_{x}^{\alpha}U_{i},\sigma_{s}\phi\right\rangle +\left\langle U_{i},q_{m}\right\rangle +\tau_{i}\Omega_{m}^{\alpha}\left\langle \partial_{x}^{\alpha}U_{i},q_{m}\right\rangle .
\end{align}
The surface integral term produced by integration by parts has been
split into known (incoming) and unknown (outgoing) parts and the boundary
condition {[}Eq. (\ref{eq:rad-boundary}){]} has been applied. Inserting
a basis function expansion for the scalar and angular flux {[}as in
Eq. (\ref{eq:basis-expansion}){]}, \begin{subequations}\label{eq:flux-expansion}
\begin{gather}
\phi=\sum_{j}V_{j}U_{j}\Phi_{j},\\
\psi=\sum_{j}V_{j}U_{j}\Psi_{j},
\end{gather}
\end{subequations}the equation becomes 
\begin{align}
 & \sum_{j}V_{j}\left[\frac{1}{c\Delta t}\left\langle U_{i},U_{j}\right\rangle +\frac{1}{c\Delta t}\tau_{i}\Omega_{m}^{\alpha}\left\langle \partial_{x}^{\alpha}U_{i},U_{j}\right\rangle +\left(U_{i},\Omega_{m}^{\alpha}n^{\alpha}U_{j}\right)_{\Omega^{\alpha}n^{\alpha}>0}-\Omega_{m}^{\alpha}\left\langle \partial_{x}^{\alpha}U_{i},U_{j}\right\rangle \right.\nonumber \\
 & \quad\left.+\tau_{i}\Omega_{m}^{\alpha}\Omega_{m}^{\beta}\left\langle \partial_{x}^{\alpha}U_{i},\partial_{x}^{\beta}U_{j}\right\rangle +\left\langle U_{i},\sigma_{t}U_{j}\right\rangle +\tau_{i}\Omega_{m}^{\alpha}\left\langle \partial_{x}^{\alpha}U_{i},\sigma_{t}U_{j}\right\rangle \right]\Psi_{m,j}\nonumber \\
 & \qquad=\sum_{j}V_{j}\left[\frac{1}{c\Delta t}\left\langle U_{i},U_{j}\right\rangle +\frac{1}{c\Delta t}\tau_{i}\Omega_{m}^{\alpha}\left\langle \partial_{x}^{\alpha}U_{i},U_{j}\right\rangle \right]\Psi_{m,j}^{n-1}+\left(U_{i},\left|\Omega_{m}^{\alpha}n^{\alpha}\right|\psi_{m}^{b}\right)_{\Omega^{\alpha}n^{\alpha}<0}\nonumber \\
 & \qquad\quad+\frac{1}{4\pi}\sum_{j}V_{j}\left[\left\langle U_{i},\sigma_{s}U_{j}\right\rangle +\tau_{i}\Omega_{m}^{\alpha}\left\langle \partial_{x}^{\alpha}U_{i},\sigma_{s}U_{j}\right\rangle \right]\Phi_{j}+\left\langle U_{i},q_{m}\right\rangle +\tau_{i}\Omega_{m}^{\alpha}\left\langle \partial_{x}^{\alpha}U_{i},q_{m}\right\rangle .\label{eq:weak-transport}
\end{align}
Note that, in general, $\Phi_{i}\neq\phi_{i}$ and $\Psi_{i}\neq\psi_{i}$.
Once Eq. (\ref{eq:weak-transport}) is solved for $\Phi$ and $\Psi$,
the solution at each point must be recovered using Eqs. (\ref{eq:flux-expansion}). 

The effect of the SUPG stabilization is to reduce oscillations and
make the system of equations easier to solve iteratively while not
affecting global particle balance \cite{bassett2019meshless}. For
the results in Sec. \ref{sec:Results}, the stabilization parameter
is set to be
\[
\tau_{i}=\frac{h_{i}}{k},
\]
where $k$ is approximately the number of points across the kernel
radius. This results in a $\tau_{i}$ that is approximately equal
to the spacing between the points. 

Because RK permits interpolation, the initial condition can be set
using initial values of the angular flux instead of needing to interpolate
coefficients, or $\Psi_{i}\bigg|_{t=0}=\psi^{\text{init}}\bigg|_{x=x_{i}}$.
This is what is done for the results in Sec. \ref{sec:Results}. 

\subsection{Methods for solution of the transport equation\label{subsec:Iterative-methods}}

The transport equation can be written in operator form as 
\begin{equation}
\mathcal{L}\Psi=\mathcal{T}\Psi^{n-1}+\mathcal{M}\mathcal{S}\Phi+r,
\end{equation}
or in terms of $\Phi$ as 
\begin{equation}
\left(\mathcal{I}-\mathcal{D}\mathcal{L}^{-1}\mathcal{M}\mathcal{S}\right)\Phi=\mathcal{D}\mathcal{L}^{-1}\left(\mathcal{T}\Psi^{n-1}+r\right),
\end{equation}
with the operators defined as 
\begin{align}
\left(\mathcal{L}\Psi\right)_{m,i}= & \sum_{j}V_{j}\left[\frac{1}{c\Delta t}\left\langle U_{i},U_{j}\right\rangle +\frac{1}{c\Delta t}\tau_{i}\Omega_{m}^{\alpha}\left\langle \partial_{x}^{\alpha}U_{i},U_{j}\right\rangle +\left(U_{i},\Omega_{m}^{\alpha}n^{\alpha}U_{j}\right)_{\Omega^{\alpha}n^{\alpha}>0}-\Omega_{m}^{\alpha}\left\langle \partial_{x}^{\alpha}U_{i},U_{j}\right\rangle \right.\nonumber \\
 & \quad\left.+\tau_{i}\Omega_{m}^{\alpha}\Omega_{m}^{\beta}\left\langle \partial_{x}^{\alpha}U_{i},\partial_{x}^{\beta}U_{j}\right\rangle +\left\langle U_{i},\sigma_{t}U_{j}\right\rangle +\tau_{i}\Omega_{m}^{\alpha}\left\langle \partial_{x}^{\alpha}U_{i},\sigma_{t}U_{j}\right\rangle \right]\Psi_{m,j},\\
\left(\mathcal{T}\Psi^{n-1}\right)_{m,i} & =\sum_{j}V_{j}\left[\frac{1}{c\Delta t}\left\langle U_{i},U_{j}\right\rangle +\frac{1}{c\Delta t}\tau_{i}\Omega_{m}^{\alpha}\left\langle \partial_{x}^{\alpha}U_{i},U_{j}\right\rangle \right]\Psi_{m,j}^{n-1},\\
\left(\mathcal{M}\mathcal{S}\Phi\right)_{i,m} & =\frac{1}{4\pi}\sum_{j}V_{j}\left[\left\langle U_{i},\sigma_{s}U_{j}\right\rangle +\tau_{i}\Omega_{m}^{\alpha}\left\langle \partial_{x}^{\alpha}U_{i},\sigma_{s}U_{j}\right\rangle \right]\Phi_{j},\\
\left(\mathcal{D}\Psi\right)_{i} & =\sum_{m}w_{m}\Psi_{m,i},\\
\left(r\right)_{i,m} & =\left(U_{i},\left|\Omega_{m}^{\alpha}n^{\alpha}\right|\psi_{m}^{b}\right)_{\Omega^{\alpha}n^{\alpha}<0}+\left\langle U_{i},q_{m}\right\rangle +\tau_{i}\Omega_{m}^{\alpha}\left\langle \partial_{x}^{\alpha}U_{i},q_{m}\right\rangle .
\end{align}
Note that using this notation, $\Phi=\mathcal{D}\Psi$. The notation
$\mathcal{L}^{-1}$ denotes the linear inverse of the $\mathcal{L}$
operator, which is block diagonal in angle. With the first-flight
source defined as 
\begin{gather}
b_{\psi}=\mathcal{L}^{-1}\left(\mathcal{T}\Psi^{n-1}+r\right),\\
b_{\phi}=\mathcal{D}b_{\psi},
\end{gather}
the equation can be simplified to 
\begin{equation}
\left(\mathcal{I}-\mathcal{D}\mathcal{L}^{-1}\mathcal{M}\mathcal{S}\right)\Phi=b_{\phi}.\label{eq:krylov-operator}
\end{equation}
Equation (\ref{eq:krylov-operator}) can be solved directly using
a matrix-free linear solver such as GMRES or iteratively using a method
such as fixed point iteration. Once the scattering source is converged,
the angular flux is recovered for use in the next time step by performing
an additional solve using the converged scalar flux,
\begin{equation}
\Psi=\mathcal{L}^{-1}\mathcal{M}\mathcal{S}\Phi+b_{\psi}.
\end{equation}
For the results in Sec. \ref{sec:Results}, the $\mathcal{L}^{-1}$
operation is performed using two packages from Trilinos \cite{heroux2005overview},
the Belos package for GMRES and the Ifpack2 package for the ILUT preconditioner.
The ILUT factorizations for each angle are precomputed at the start
of the time step and reused to minimize computation. The iterations
to converge the scattering source {[}the solution of Eq. (\ref{eq:krylov-operator}){]}
are also performed using GMRES from Belos without a preconditioner.
For a discussion on preconditioners for the scattering iterations,
see Sec. \ref{sec:Conclusions}. 

\subsection{Refinement of the angular quadrature\label{subsec:Angular-refinement}}

For the problems in Sec. \ref{sec:Results}, the Gauss-Legendre quadrature
is used in 1D, while the LDFE (linear discontinuous finite element)
quadrature \cite{jarrell2011discrete} is used in 2D and 3D. The LDFE
quadrature is hierarchal, meaning that each octant of the unit sphere
can be further subdivided into four ordinates, which can themselves
be subdivided and so on. This can be used to produce a high density
of angular ordinates in chosen directions object to prevent ray effects,
which as shown in Sec. \ref{subsec:Asteroid}. Given a goal quadrature,
the refined angular discretization keeps all ordinates from the goal
quadrature that hit the object and combines the ordinates that do
not hit the object inasmuch as is possible (Alg. \ref{alg:angular-refinement}).
This significantly reduces the number of angles needed for a given
number of rays from a small source to hit a distant object. 

\section{Results\label{sec:Results}}

In this section, five problems are considered to test the discretization
described in Sec. \ref{sec:Transport} with the integration in Sec.
\ref{sec:Integration}. The first two problems use the method of manufactured
solutions. The third problem considers a purely absorbing medium,
while the fourth problem considers a purely scattering medium. The
final problem shows a possible application of the code to simulate
an asteroid absorbing a large quantity of radiation from a distant
source. All the problems use a kernel sampling radius of 4 neighbors
(i.e., the equivalent smoothing scale is 4 times the local particle
spacing) and an RK order of one, as described in Sec. \ref{sec:Integration},
except for the purely absorbing problem, which also explores other
combinations of RK order and neighbors. Note that for uniformly spaced
points, a sampling radius of 4 neighbors implies a total number of
overlap neighbors for each point of 8 (1D), 50 (2D), and 268 (3D).

The first two sections use the relative error as a measure for convergence.
This is defined as 
\begin{equation}
\epsilon_{rel}=\dfrac{\sum_{i}\left|\phi_{i}^{\text{numeric}}-\phi_{i}^{\text{analytic}}\right|}{\sum_{i}\phi_{i}^{\text{analytic}}}\label{eq:relative-err}
\end{equation}
for the numeric and analytic scalar fluxes at the MLPG centers $i$. 

\subsection{Manufactured problems}

The method of manufactured solutions works by selecting a solution
for $\psi$, solving for a source $q$ by inserting this solution
into the continuous transport equation {[}Eq. (\ref{eq:transport-equation}){]},
assigning the boundary source $\psi_{b}$ to be equal to the solution,
and then calculating a numerical solution using the discretized transport
equation {[}Eq. (\ref{eq:weak-transport}){]} for comparison to the
original solution. 

The spatial and time convergence of the manufactured problems is considered
in 1D, 2D, and 3D. In 1D, due to the low cost of integration and because
the integration cells are not subdivided, the integration is performed
using a 64-point Gauss-Legendre quadrature. In 2D, the integration
of the subcell triangles is performed using a tenth-order symmetric
quadrature with 25 points, while in 3D, the integration of the surface
triangles and the subcell tetrahedra is performed using a third-order
symmetric quadrature with 8 points. For more information on the integration
quadratures, see Sec. \ref{sec:Integration}. 

The spatially-dependent results are run for several cases between
$8^{d}$ and $128^{d}$ points (for the dimension $d$) without time
dependence. For the time-dependent case, the simulation is run until
$t=1$ with time steps between 0.001 and 1.0 and $64^{d}$ points.
The minimum time step is increased to 0.01 in 2D and 0.1 in 3D. In
each case the points are laid down in a spatially uniform lattice
configuration. For the non-uniform cases, the point positions are
randomly perturbed by up to 0.2 times the point distance in each dimension,
or $x^{\alpha}=x^{\alpha}\pm\Delta x^{\alpha}\gamma^{\alpha}$, where
$-0.2\leq\gamma^{\alpha}\leq0.2$ is randomly generated for each point
and dimension independently and $\Delta x^{\alpha}$ is the point
spacing for the given dimension. 

\subsubsection{Sinusoidal manufactured problem}

The first manufactured solution,

\begin{equation}
\psi_{\text{sinusoidal}}=1+\frac{1}{2\pi}\prod_{\alpha}\cos\left(\pi\left(x^{\alpha}+t\right)\right),
\end{equation}
is designed to test convergence of the discretized transport equation
{[}Eq. (\ref{eq:weak-transport}){]}. The solution is chosen such
that the manufactured source never becomes negative, which would be
unphysical. To ensure that the integration of the cross sections {[}Sec.
\ref{sec:Integration}{]} works correctly, the scattering and absorption
opacities are also chosen to have sinusoidal values that are out of
phase with the solution and one another,
\begin{gather}
\sigma_{a}=1+\frac{2}{3}\prod_{\alpha}\cos\left(3x^{\alpha}\right),\\
\sigma_{s}=1+\frac{3}{4}\prod_{\alpha}\sin\left(2x^{\alpha}\right).
\end{gather}
The domain is $-1\leq x^{\alpha}\leq1$. For the steady-state case,
the manufactured solution is fixed at $t=0$.

\begin{figure}[H]
\begin{centering}
\subfloat[Manufactured sinusoidal solution, uniform.]{\begin{centering}
\includegraphics[width=0.46\textwidth,trim={0cm 0cm 0cm 0cm},clip]{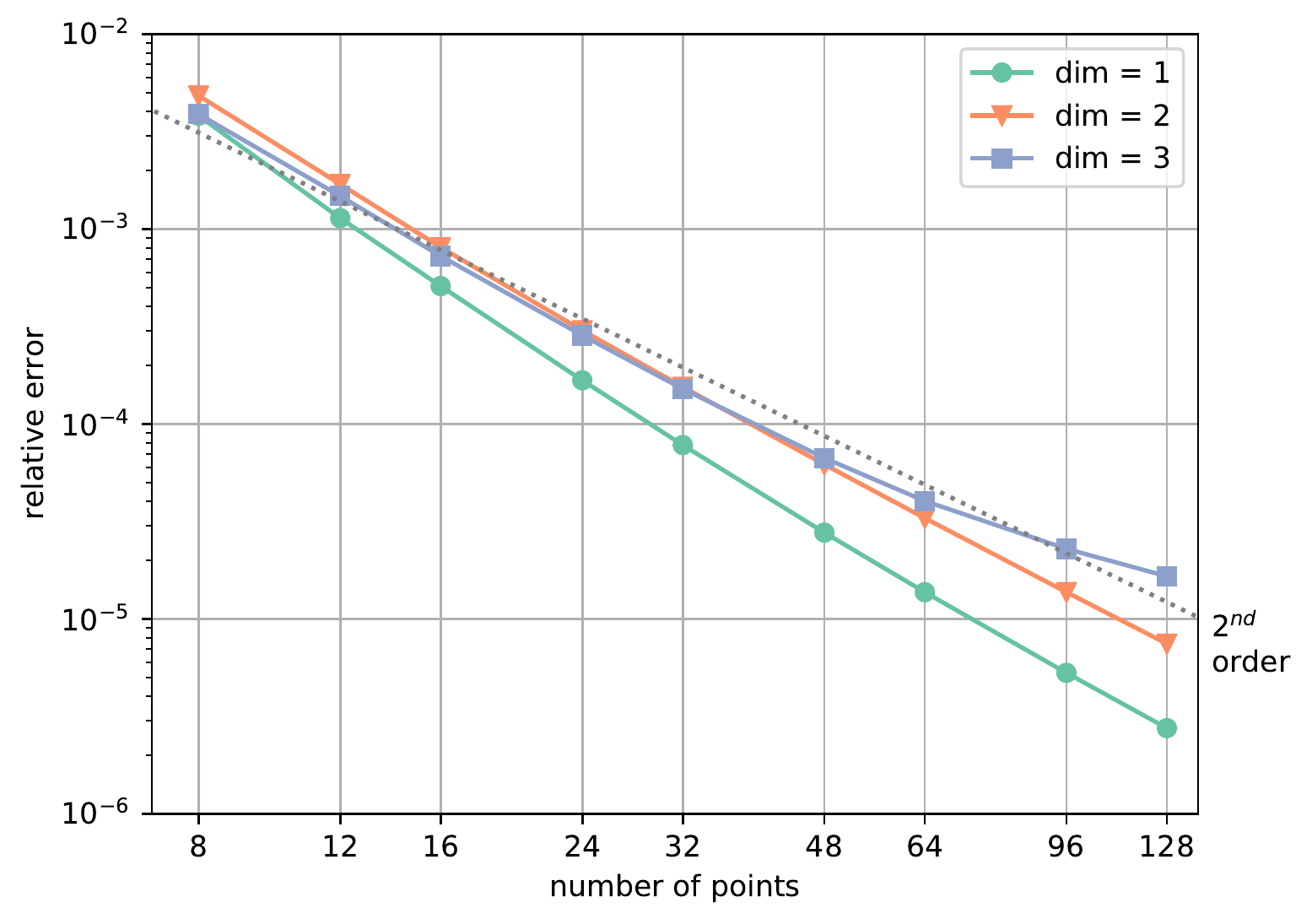}
\par\end{centering}
\label{fig:man-sin-steady}}\hfill{}\subfloat[Manufactured wave solution, uniform.]{\begin{centering}
\includegraphics[width=0.46\textwidth,trim={0cm 0cm 0cm 0cm},clip]{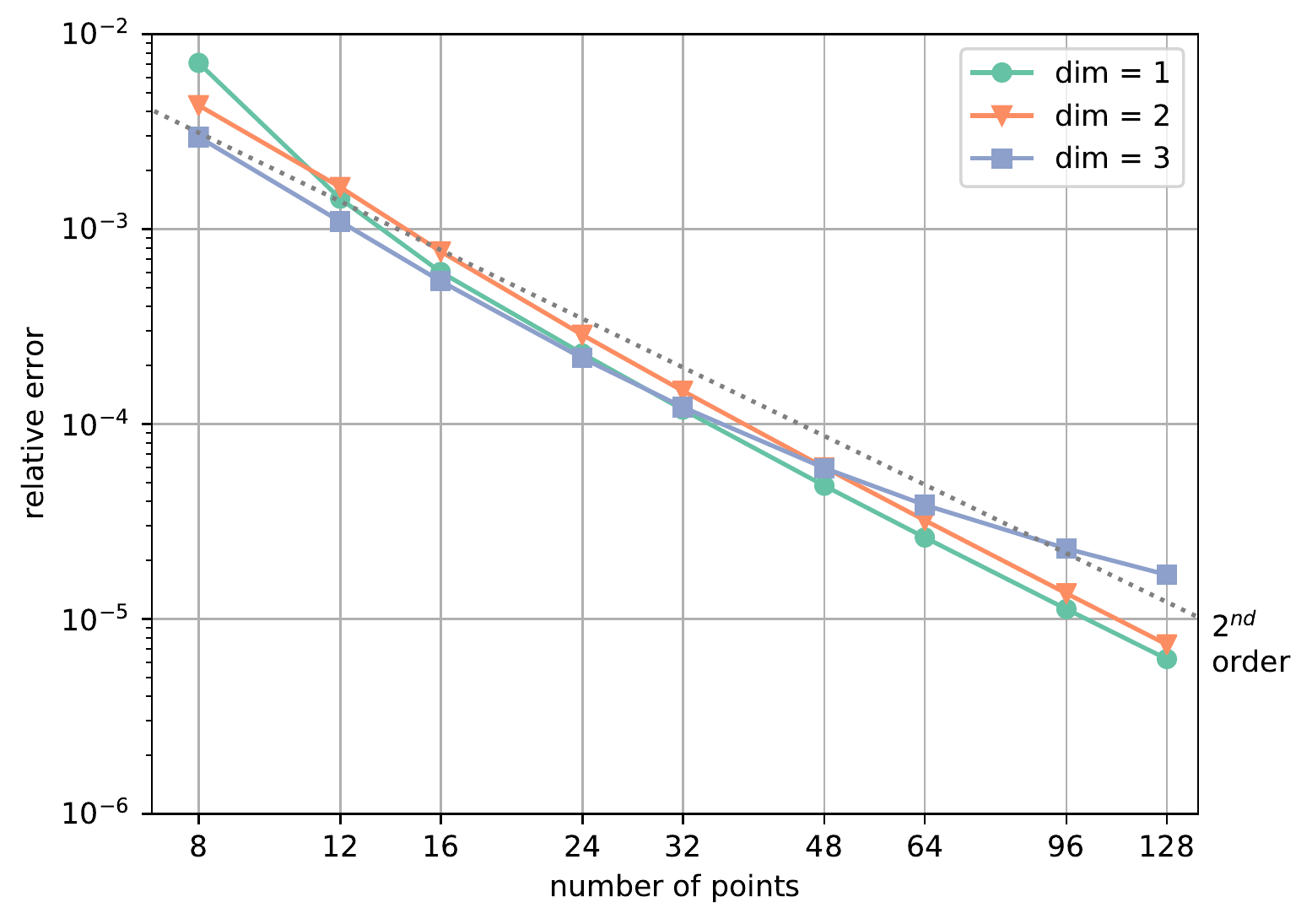}
\par\end{centering}
\label{fig:man-wave-steady}}
\par\end{centering}
\begin{centering}
\subfloat[Manufactured sinusoidal solution, perturbed.]{\begin{centering}
\includegraphics[width=0.46\textwidth,trim={0cm 0cm 0cm 0cm},clip]{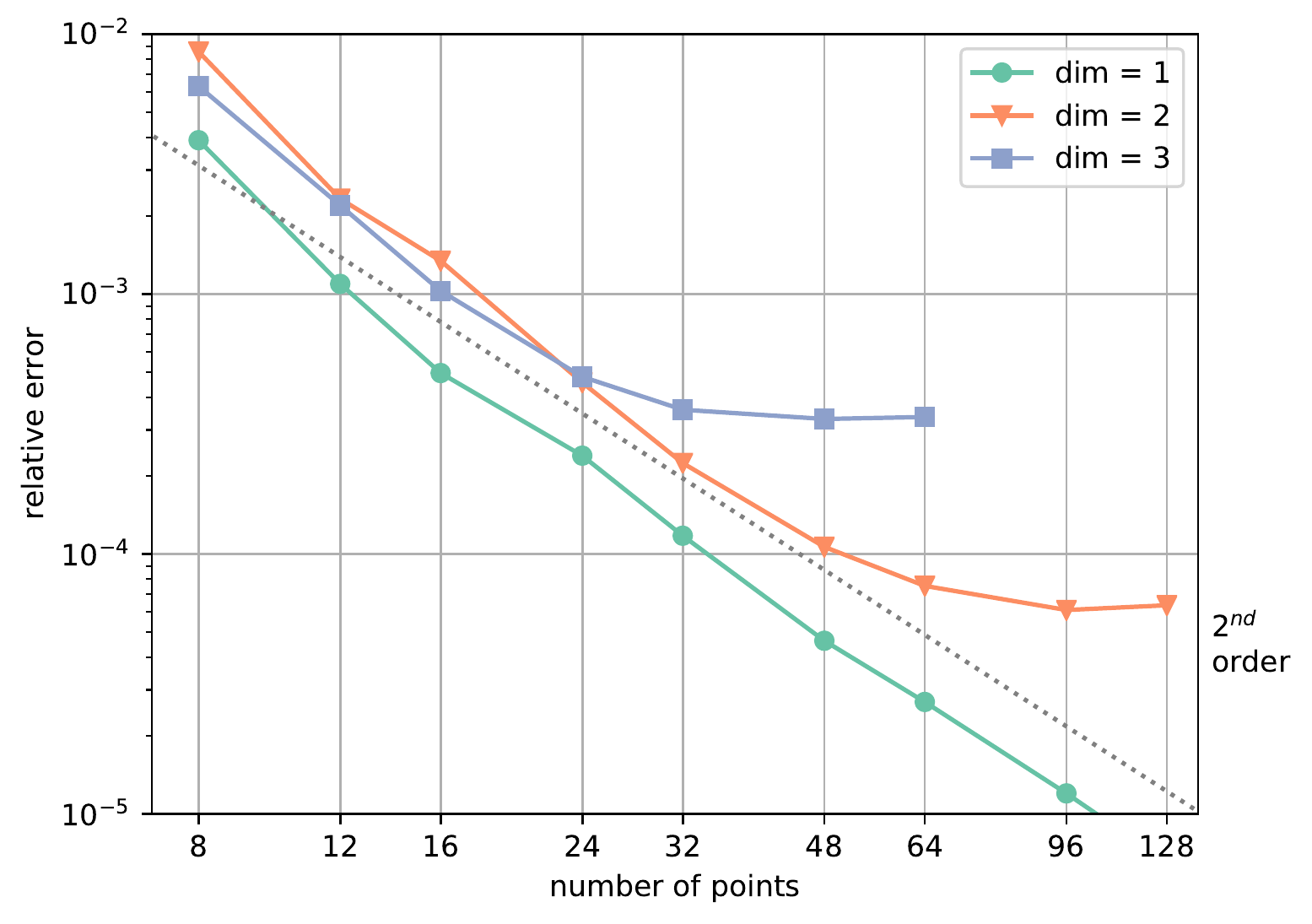}
\par\end{centering}
\label{fig:man-sin-steady-perturb}}\hfill{}\subfloat[Manufactured wave solution, perturbed.]{\begin{centering}
\includegraphics[width=0.46\textwidth,trim={0cm 0cm 0cm 0cm},clip]{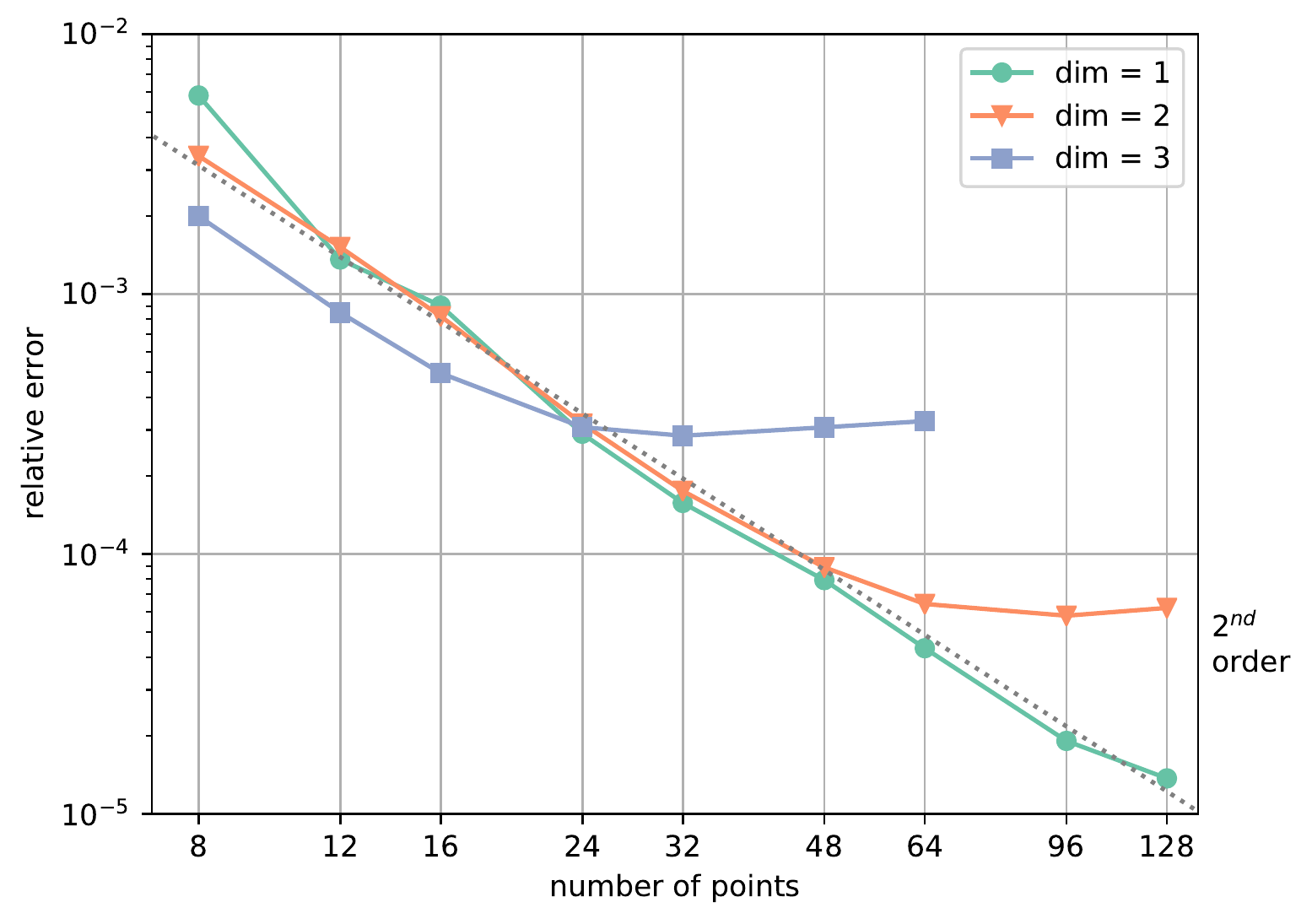}
\par\end{centering}
\label{fig:man-wave-steady-perturb}}
\par\end{centering}
\caption{Spatial convergence of the manufactured solutions at steady-state
in 1D, 2D, and 3D, with the dotted line indicating second-order convergence.
For two of the cases, the point positions are randomly perturbed.
The perturbation may cause insufficient support for the RK functions
or lower integration accuracy, leading to a lack of convergence. }
\end{figure}

The spatial convergence results in Fig. \ref{fig:man-sin-steady}
indicate second-order convergence in 1D, 2D, and 3D, as expected for
linear RK corrections. The 3D results eventually plateau around 96
points. It is likely that the difference between the numeric and analytic
solution is reaching the accuracy limit of the third-order quadrature
in 3D, which appears to reduce the convergence order to first-order.
The inclusion of spatially-dependent cross sections does not appear
to hinder convergence. 

When the point positions are randomly perturbed (Fig. \ref{fig:man-sin-steady-perturb}),
the convergence rate stays the same, with a caveat. The algorithm
for calculating kernel extents is designed for hydrodynamics and requires
that the average level of support is above a certain threshold, not
the support for each point. It is possible that in 2D and 3D, the
solution reaches the accuracy of the poorly-supported RK calculation
and stops converging for certain points. Before this occurs, the convergence
is second-order. When the calculation is run with a higher kernel
extent (not pictured), the quantitative behavior is similar but the
error levels off at a lower value. The other difference between dimensions
is the integration quadrature, which is coarser and less accurate
between dimensions. This could be contributing to the leveling off
of the error. 

\begin{figure}[H]
\begin{centering}
\subfloat[Manufactured sinusoidal solution.]{\begin{centering}
\includegraphics[width=0.46\textwidth,trim={0cm 0cm 0cm 0cm},clip]{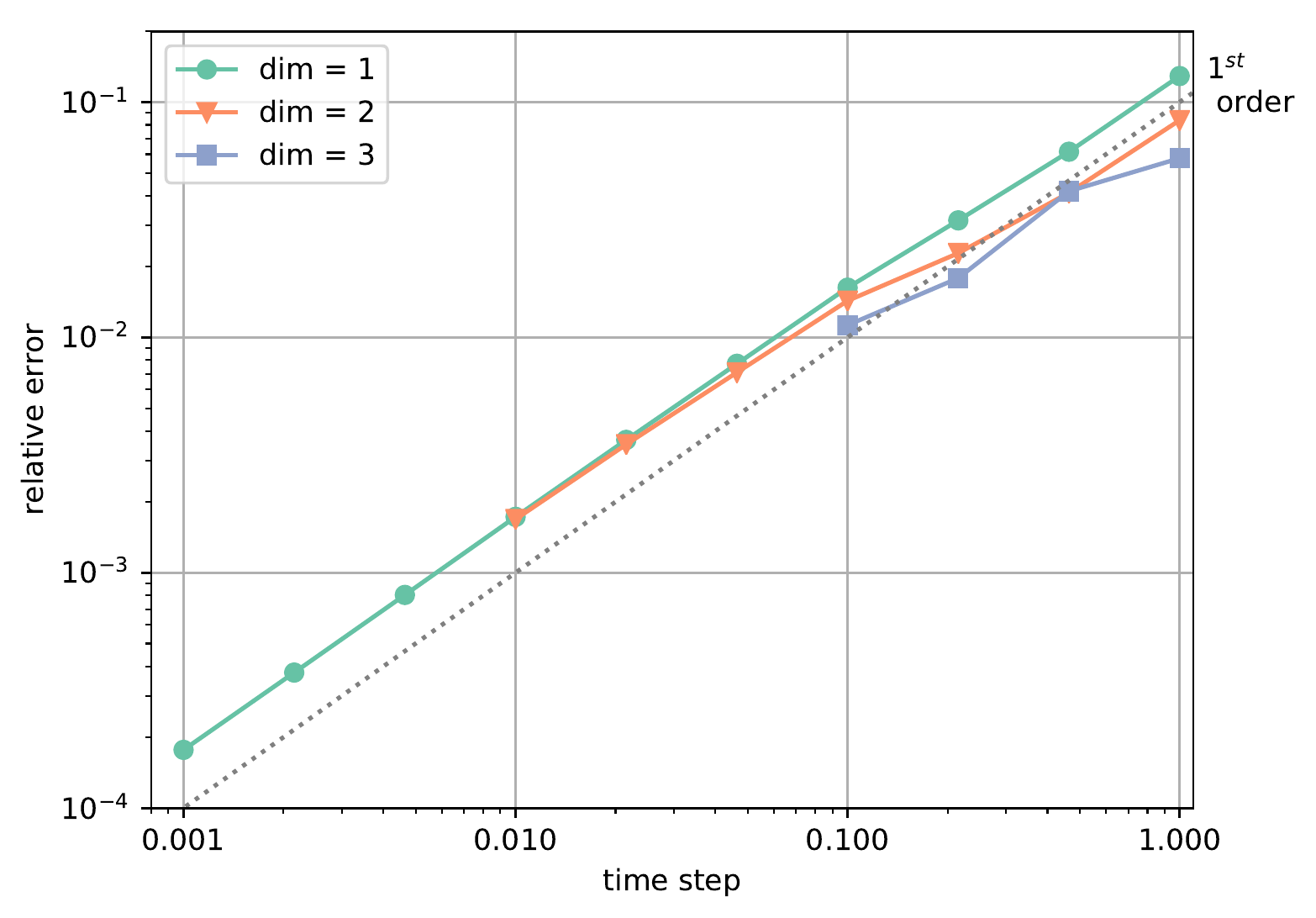}
\par\end{centering}
\label{fig:man-sin-time}}\hfill{}\subfloat[Manufactured wave solution.]{\begin{centering}
\includegraphics[width=0.46\textwidth,trim={0cm 0cm 0cm 0cm},clip]{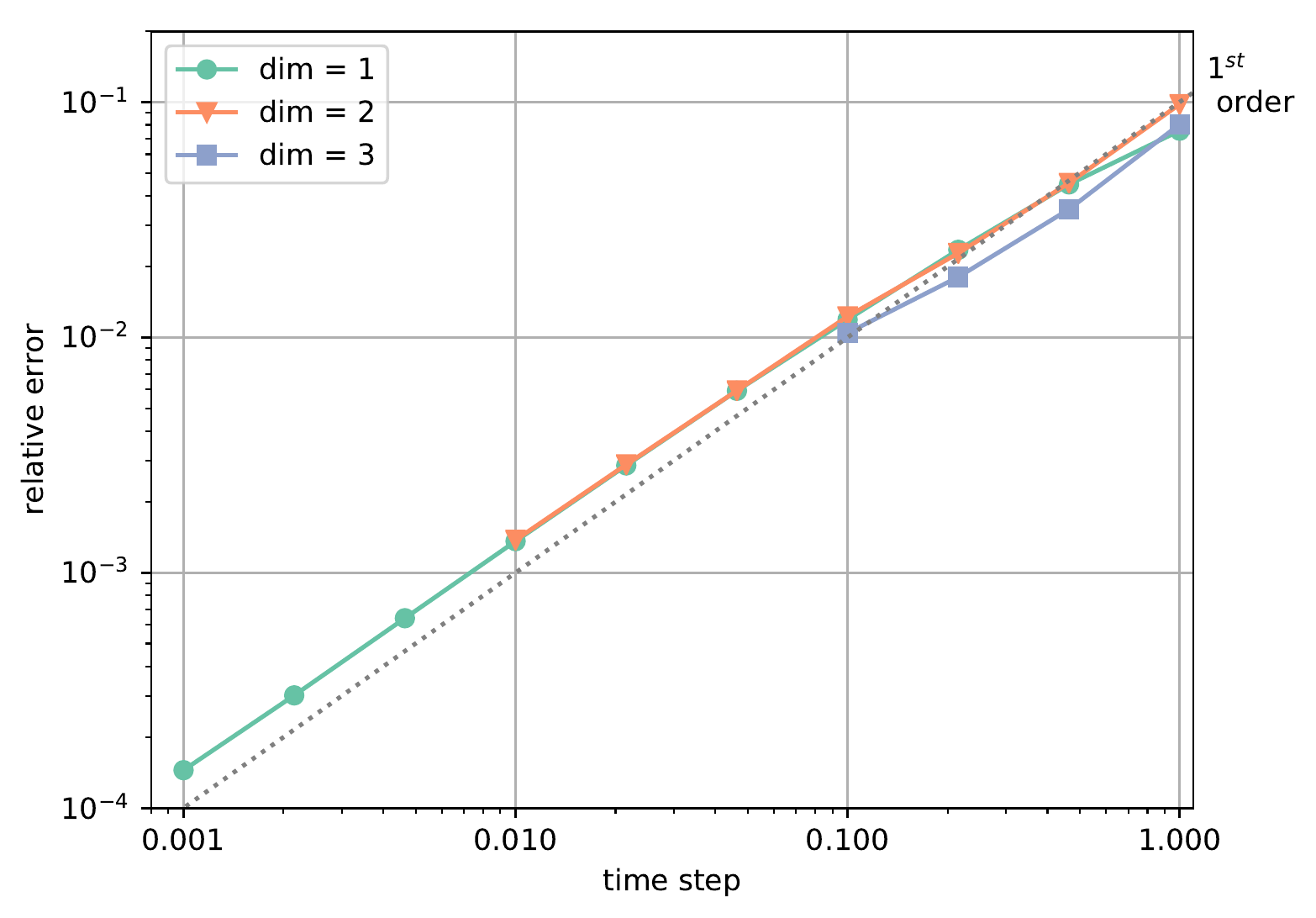}
\par\end{centering}
\label{fig:man-wave-time}}
\par\end{centering}
\caption{Temporal convergence of the manufactured solutions in 1D, 2D, and
3D, with the dotted line indicating first-order convergence.}
\end{figure}

The temporal convergence rate is first-order (Fig. \ref{fig:man-sin-time}),
as expected from the backward Euler time discretization. The 2D and
3D results have similar or lower error than the 1D results for a similar
number of points, which may be due to the higher level of connectivity
in 2D and 3D. For this problem, the time discretization error even
with the smallest time step considered (0.001) is similar to the spatial
discretization error with $32^{d}$ points. It is expected that to
increase the accuracy of a time-dependent simulation at the point
where the temporal and spatial discretization errors are similar,
the time step would need to be decreased as the distance between points
squared. 

\subsubsection{Outgoing wave manufactured problem}

The second manufactured problem represents a wave traveling from the
origin outward, 

\[
\psi_{\text{wave}}=1+\frac{1}{t^{2}+6}\exp\left(-10\left[\left|x\right|-t\right]^{2}\right),
\]
with constant opacities of $\sigma_{a}=0.5$ and $\sigma_{s}=2.0$
and a domain of $-1\leq x^{\alpha}\leq1$. For the steady-state case,
the manufactured solution is fixed at $t=0.5$. 

As in the sinusoidal case, the spatial convergence is second-order
(Fig. \ref{fig:man-wave-steady}) and the temporal convergence is
first-order (Fig. \ref{fig:man-wave-time}). The magnitude of the
error is similar in the wave and the sinusoidal case, and just as
in the sinusoidal case, the error in 3D plateaus on the spatial convergence
plot, probably due to the integration error. The perturbed version
of the steady-state problem (Fig. \ref{fig:man-wave-steady-perturb})
has similar behavior to the sinusoidal case described above, with
second-order convergence until reaching issues with either RK kernel
support or integration. 

\subsection{Purely absorbing problem\label{subsec:Purely-absorbing}}

One challenge in transport is handling highly absorptive regions without
incurring negative fluxes. In this problem, a single ray with $\Omega=\left\{ 1,0,0\right\} $
is incident on a purely absorbing slab with a domain $0\leq x\leq1$,
which is modeled in 1D. First, a constant cross section of $\sigma_{a}=5$
is considered with a variable number of points between 8 and 64. Then,
the number of points is held constant at $32$ and the cross section
is varied between $1$ and $64$. The points are again placed uniformly. 

Convergence results are shown in Fig. \ref{fig:purely-absorbing-err}
for a few cases of the RK order and the number of neighbors, which
is the number of other points across a kernel radius for the base
connectivity used to create the overlap connectivity. The number of
neighbors for the reduced-radius kernels should be at least one higher
than the RK order to prevent the system in Eq. (\ref{eq:rk-matrix})
from being singular, which means that the number of neighbors for
the original kernels (which is the number reported here) should be
two times the RK order plus one. For zeroth-order RK corrections,
the solution converges with approximately second-order accuracy. For
first-order corrections, the solution converges with approximately
second-order accuracy for 4 neighbors and between second and third
order for 6 neighbors. With second-order corrections and 6 neighbors,
the convergence order is between third and fourth. In general, for
a smooth solution, the expected convergence order is one greater than
the RK order. 

Results for the case with a constant number of points and a changing
cross section are shown in Fig. \ref{fig:purely-absorbing-xs}. For
this problem, in which the primary gradient is at the edge of the
problem with the lowest point density, the MLPG approach requires
around one point per mean free path of the material to avoid negativities,
which is reflected in the results. The solution begins to exhibit
negativities at $\sigma_{a}=32$ for the case with 6 neighbors, while
for 4 neighbors, the negativities show up for $\sigma_{a}=64$ and
above. Note that the error is an absolute error, since the normalization
would otherwise skew the results. As in the convergence study, an
increasing number of neighbors and RK order decrease the solution
error. 

Based on these results, it may be tempting to use kernels with large
radii and high RK order for other problems to increase solution accuracy.
One issue with this is computational cost, which increases significantly
in 2D and 3D as the function radii increase. Since the RK order is
limited for kernels with small radii, this also limits the RK order.
For instance, in 3D, the number of neighbors increases as $r^{3}$,
where $r$ is the kernel radius, so moving from 4 neighbors across
the kernel radius to 6 will more than triple the cost. This also increases
the difficulty of solving the transport system (the $\mathcal{L}^{-1}$
operation). Another issue is negativities, which are present for all
the 32-point simulations with 6 neighbors but not for any 32-point
simulations with 4 neighbors. These negativities can become amplified
in time-dependent problems, where a negative absorption becomes an
unphysical source of particles. For more discussion on negative fluxes,
see Sec. \ref{sec:Conclusions}. 

\subsection{Crooked pipe problem}

This problem is described in Ref. \cite{smedley-stevenson2015benchmark}
as a test of diffusion synthetic acceleration (DSA). Results for the
MLPG code with the Krylov iteration as described in Sec. \ref{subsec:Iterative-methods}
are compared to those from a code based on the discontinuous finite
element method (DFEM) with acceleration based on a variable Eddington
factor (VEF), as described in Ref. \cite{yee2020quadratic}. 

The geometry for the problem is shown in Fig. \ref{fig:crooked-geometry},
with $\sigma_{s}=200$ in the wall and $\sigma_{s}=0.2$ in the pipe.
The absorption cross section is zero in both regions. The relatively
coarse angular discretization with 16 ordinates is identical between
the MLPG and DFEM codes. The MLPG results have a spatial discretization
with 716,800 equally-spaced points (or 160 by 160 points per unit
area), while the DFEM results are calculated on a mesh with 1,335,296
elements, with a higher density of elements placed near the pipe-wall
boundary. The MLPG points and DFEM mesh at quarter resolution (or
16 times fewer points) is shown in Fig. \ref{fig:crooked-mesh}. The
units for the problem are set such that the speed of light is $c=1$.
The problem is run until $t=20$ with a fixed time step of $\Delta t=0.1$. 

\begin{figure}[H]
\begin{centering}
\subfloat[$t=10$]{\begin{centering}
\includegraphics[width=1.0\textwidth,trim={5cm 4cm 7cm 2cm},clip]{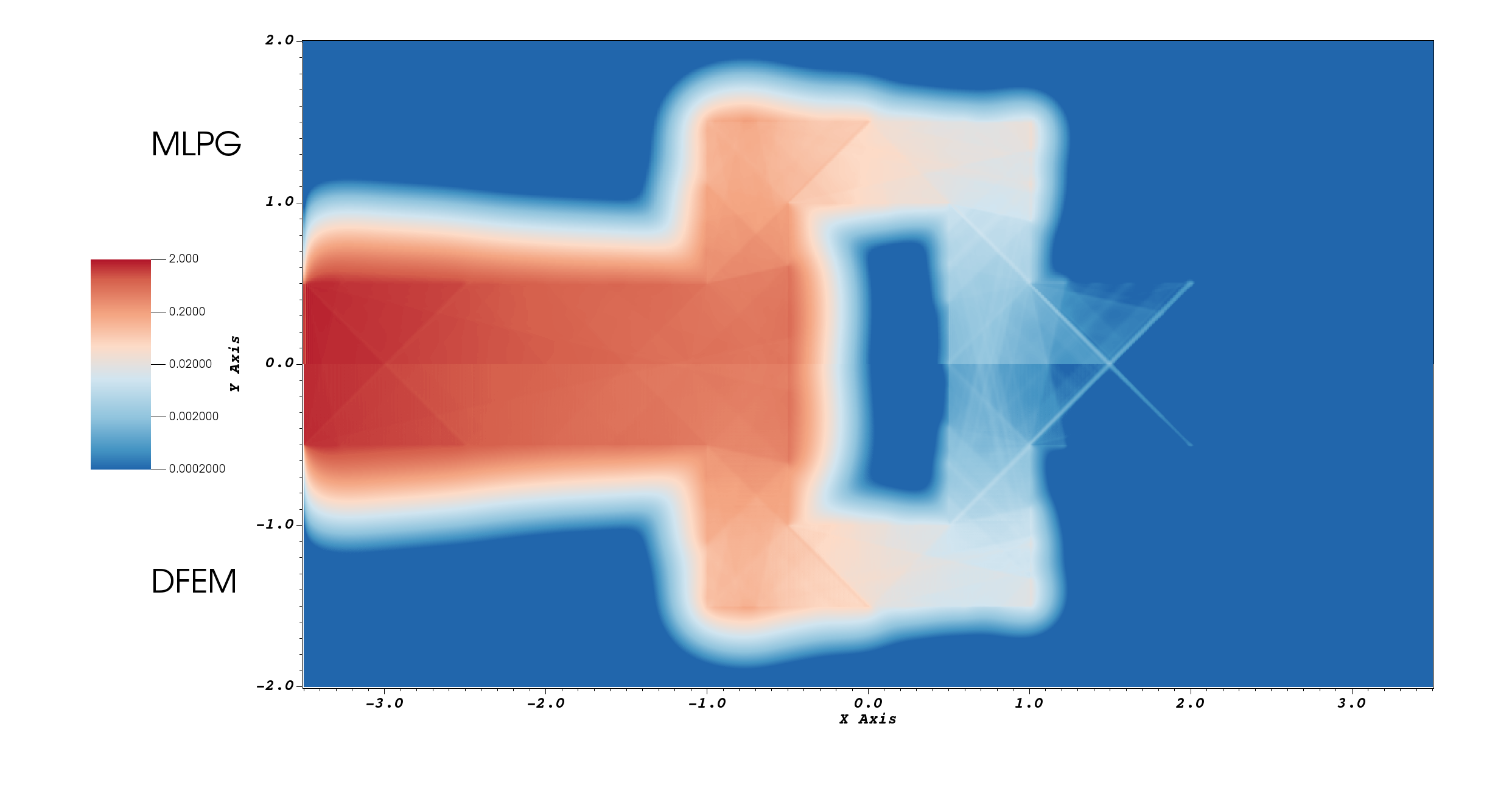}
\par\end{centering}
\centering{}}
\par\end{centering}
\begin{centering}
\subfloat[$t=20$]{\begin{centering}
\includegraphics[width=1.0\textwidth,trim={5cm 4cm 7cm 2cm},clip]{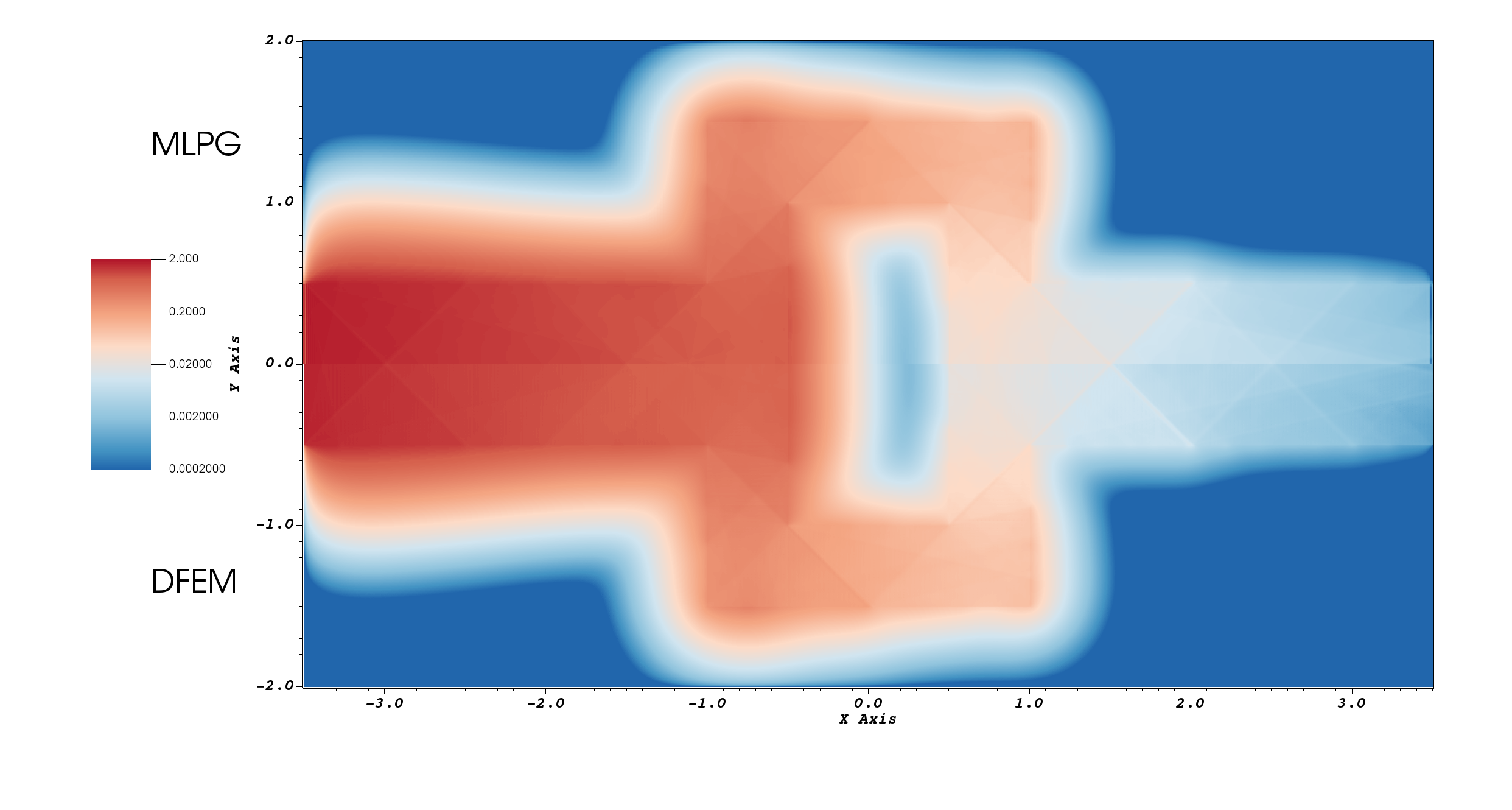}
\par\end{centering}
}
\par\end{centering}
\caption{The scalar flux for the MLPG (top) and DFEM (bottom) solutions to
the crooked pipe problem at two times.}
\label{fig:crooked-comparison}
\end{figure}

A comparison of the crooked pipe results at $t=10$ and $t=20$ is
shown in Fig. \ref{fig:crooked-comparison}. The propagation speed
of the radiation appears is nearly identical between the two codes.
For the $t=10$ plot, the radiation would have traveled 10 unit distance
at most in the 10 unit time (since $c=1$). The minimum path the radiation
could take to reach the plane at $x=2$ from the source at $x=-3.5$
is $6.5$ unit distance. Depending on the direction, the actual distance
the radiation would need to travel to reach the plane $x=2$ would
be at between 7 and 12 unit distance. The strongest visible ray is
at $\Omega^{\alpha}=1/\sqrt{3}$, which would have reached $x=2$
at $t=9.5$ if scattering and corners were neglected. As the radiation
appears to have just reached $x=2$ at $t=10$, the calculated time
of arrival is close to the distance the radiation would have traveled
in that time. 

The results show significant ray effects, but because the two codes
use the same angular quadrature, the effects appear to be the same.
Before reaching the crooked part of the problem, there are no significant
differences visible between the two solutions. After the radiation
has gone around the obstacle, the MLPG solution is higher in magnitude,
which is visible in the $t=10$ plot near the right edge of the obstacle
or in the $t=20$ plot at the exiting surface of the pipe. Part of
this could be due to the higher resolution along the pipe-wall interface
in the DFEM simulation, which could affect the scattering rate at
the interface. The contours of the solution near the interfaces also
line up very closely, except at the wall edge at $x=0.5$, where the
MLPG solution reaches a further through the wall, which could again
be due to the lower resolution near the interface. 

As mentioned before, this problem has been used as a test of DSA.
The Krylov solution procedure {[}for the solution of the MLPG system
in Eq. (\ref{eq:krylov-operator}){]} works for this case without
preconditioning, with an average of 64 iterations to converge. As
the ILUT factorization of the matrices representing the $\mathcal{L}^{-1}$
operation is performed once and stored, this doesn't increase the
total simulation time by nearly 64 times more than a single iteration,
as the factorization is a far larger cost than a single solve. Within
each scattering iteration, the transport GMRES solver converges to
a tolerance of $10^{-15}$ in 55 iterations on average. The DFEM solution,
however, required only one transport solve and two VEF solves per
time step, which if applied to the MLPG solution could open up more
cost-effective methods for solving for the scattering source. 

For general reference, the RK transport code takes 15,943 seconds,
or 79 seconds per time step, to run the crooked pipe problem with
16 ordinates and 716,800 points on 288 processors, which equates to
2,488 points or 39,808 unknowns per processor on average. This includes
the time for integration, computation of the ILUT preconditioners
for all 16 directions, and convergence of the solution and scattering
source. With an effective preconditioner and possibly avoiding ILUT
decompositions, the solve time would decrease significantly. The need
for appropriate preconditioning is discussed further in Sec. \ref{sec:Conclusions}. 

\subsection{Asteroid problem\label{subsec:Asteroid}}

One motivation for combining radiation transport with a smoothed particle
hydrodynamics code is for a planetary defense application: the deflection
of an asteroid due to radiation from a standoff nuclear burst. In
this scenario, the absorbed radiation energy ablates the surface of
the asteroid, causing material to blow off and alter the orbit of
the asteroid via momentum conservation. This problem is a simplified
version of that scenario, a spherical rock ``asteroid'' that absorbs
radiation from a distant point source in 2D. This problem is similar
to the purely-absorbing version of Kobayashi benchmarks \cite{kobayashi20013d},
which also have features that are difficult to angularly resolve and
a small source emitting particles into a void. The asteroid has an
absorption cross section of $\sigma_{a}=10.0$, while the medium surrounding
the asteroid has an absorption cross section of $\sigma_{a}=0.001$.
Neither material includes scattering. The asteroid has a radius of
35 and is centered at the origin. The point source is located a distance
of 70 away from the surface of the asteroid. The asteroid can be modeled
by a shell, since almost all of the radiation is absorbed at the surface
of the asteroid. 

\begin{figure}[H]
\begin{centering}
\includegraphics[width=0.85\textwidth,trim={15cm 5cm 7cm 3cm},clip]{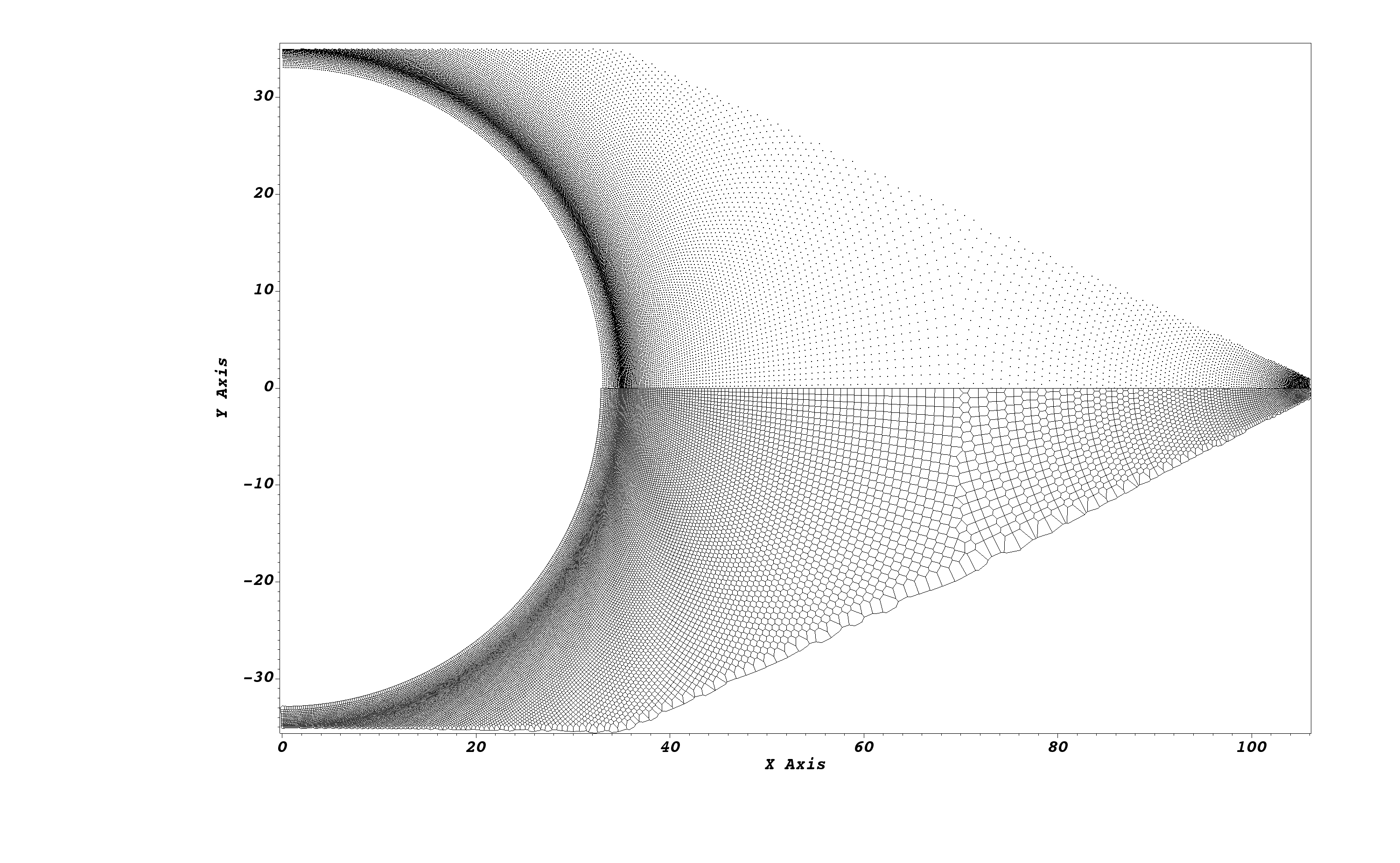}
\par\end{centering}
\caption{The solution points (top) and integration mesh (bottom) for the asteroid
problem. Note that the resolution of the integration mesh mirrors
that of the kernels. The integration cells are further subdivided
into triangles for integration.}
\label{fig:asteroid-geometry}
\end{figure}

The shell of the asteroid is set to be 20 mean free paths thick. The
distance between points is set to be 0.2 at the inside of the shell
of the asteroid, 0.1 at the outside of the shell, 1.0 halfway between
the source and the asteroid, and 0.1 near the source. The initial
angular quadrature of 4,096 ordinates is refined as described in Sec.
\ref{subsec:Angular-refinement} down to 550 ordinates, of which 480
hit the asteroid. The problem is run with a single time step large
enough for the radiation to propagate throughout the domain. Afterward,
the numeric solution is compared to the analytic solution, which is
derived in App. \ref{sec:Analytic-asteroid}. The solution points
and integration mesh for this problem are shown in Fig. \ref{fig:asteroid-geometry}.
Note that the integration mesh is further broken down into triangles
for use with standard quadratures (Sec. \ref{sec:Integration}). 

\begin{figure}[H]
\begin{centering}
\includegraphics[width=0.85\textwidth,trim={5cm 6cm 7cm 9cm},clip]{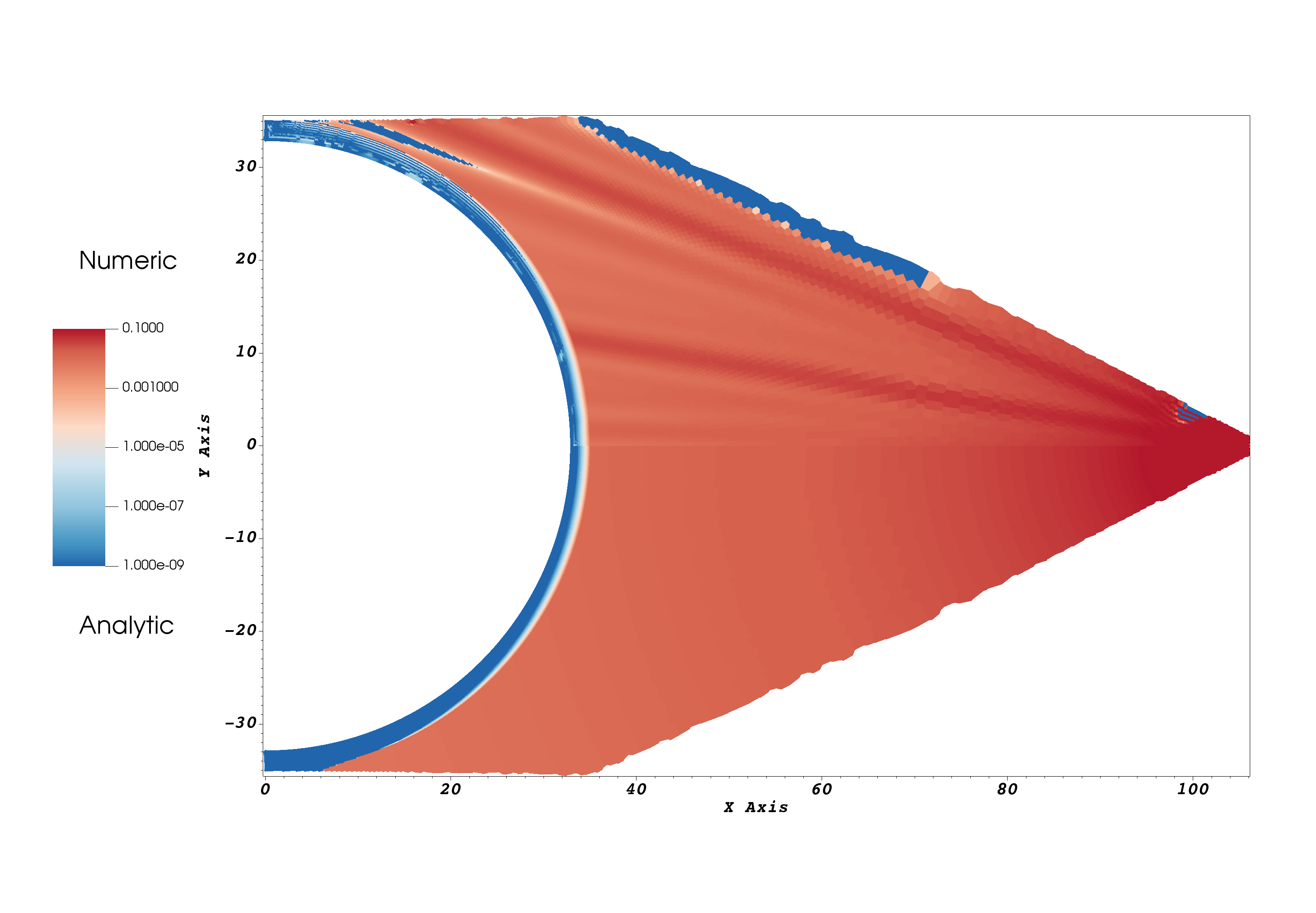}
\par\end{centering}
\caption{The numeric (top) and analytic (bottom) scalar flux solution to the
asteroid problem. The low resolution for rays that miss the asteroid
is by design. Note the ray effects where the radiation propagation
direction is parallel to the surface of the asteroid. }
\label{fig:asteroid-solution}
\end{figure}

The analytic and numeric solutions to the asteroid problem are shown
in Fig. \ref{fig:asteroid-solution}. The most obvious difference
at first glance is the large areas at the top of the numeric solution
where the solution is close to zero. These are areas where, by design,
the angular refinement has not put a sufficient number of angles to
resolve the solution. These should not affect the solution at the
asteroid, since there, the solution should be sufficiently resolved
in the angular domain. While there are 480 rays that hit the asteroid
from the point source, the radiation is still not angularly uniform
in the region that is resolved, as can be seen by the more intense
ray that hits the asteroid around the point $\left\{ 11,33\right\} $. 

Near the $y=0$ plane, where there is approximately one point per
mean free path in the direction the solution is changing, the contours
of the analytic and numeric solutions line up very well, with the
exception of the aforementioned oscillations. This agrees with the
purely absorbing results (Sec. \ref{subsec:Purely-absorbing}), in
which the solutions with around one point per mean free path showed
higher accuracy and few oscillations compared to those with more than
one point per mean free path. 

There are two connected difficulties in this problem, which are ray
effects and negativities. Oscillations can be seen toward the inner
surface of the asteroid at all positions, but the oscillations are
by far the worst where the radiation is traveling nearly parallel
to the surface of the asteroid. At these points, the solution will
change from the vacuum solution just outside of the surface to nearly
zero inside of the surface. This causes oscillations that lead to
negativities. The SUPG stabilization does a good job of handling the
oscillations that may develop in the direction of radiation propagation
(near the $y=0$ plane), but more consideration is needed to prevent
the oscillations that develop perpendicular to the radiation propagation
direction or when the solution changes discontinuously. 

This problem is designed to stress the code and show opportunities
for future work. If the results needed to be accurate, the source
could be analytically calculated just before it hits the asteroid
and inserted as a boundary source there, which would reduce much of
the need for a refined and specialized quadrature. The negatives,
however, would persist, which is something that would need to be addressed
before this calculation would work well in a time-dependent or thermal
radiative transfer scenario, as discussed in Sec. \ref{sec:Conclusions}. 

\section{Conclusions and future work\label{sec:Conclusions}}

The MLPG discretization in this paper simplifies the process running
a problem with meshless transport. The fully-implicit time differencing
is stable for large time steps. The SUPG stabilization works to prevent
oscillations and increase the efficiency of inverting the transport
matrix. The integration with a Voronoi diagram is robust and follows
the resolution of the meshless points without user input. The SUPG
stabilization and Voronoi integration add complexity to the code,
which could be a barrier to entry, but reduces the need for specialized
solvers or repeated adjustments of a background mesh. 

The RK functions used in the discretization permit higher-than-second-order
convergence, but practically, the radii of the kernels should often
be minimized to reduce negativities and computation cost, which constrains
the RK order. With the fully-implicit time discretization and first-order
RK corrections, the results are consistent with second-order convergence
in space and first-order convergence in time. For a purely absorbing
problem with an incoming source, higher-order convergence is achieved
in space with second-order RK corrections and larger kernel radii.
The two manufactured solutions show the capability to represent spatially-dependent
cross sections and converge in 1D, 2D, and 3D. 

There are at least three issues that remain to be resolved. The first
is negativities. In the purely-absorbing problem, around one point
is needed per mean free path to avoid negativities. In the asteroid
problem, negativities are difficult to avoid due to ray effects, as
the SUPG stabilization applies numerical diffusion only in the direction
of radiation propagation. A negative flux fixup method such as the
zero-and-rescale approach \cite{hamilton2009negative} may work for
meshless transport, but care would need to be taken to rescale a quantity
that should always be positive and not the expansion coefficients.
While this could inhibit the effects of oscillations on time-dependent
problems and perhaps keep them from growing, it would be much more
difficult to remove oscillations entirely. 

The second issue is preconditioning. In the crooked pipe problem,
the solution converged when using only GMRES to converge the scattering
source and agreed well with a DFEM solution, but this required many
iterations to achieve. Combining the Krylov solve with a method such
as DSA \cite{warsa2004krylov} could significantly reduce the number
of iterations needed to converge the scattering source. The MLPG transport
equation without SUPG should have the diffusion limit, similar to
a high-order DFEM discretization \cite{haut2018dsa}, but it is not
apparent whether the same is true with SUPG. The process of deriving
DSA for the SUPG system should give information on whether it has
the diffusion limit and if not, what changes may be made to the discretized
system to ensure it has the diffusion limit. 

The third issue is ensuring proper support for the RK kernels. As
shown in the manufactured problems, randomly perturbing the point
positions can lead to a limit on convergence. Ensuring that every
point has the needed support individually through a more robust calculation
may resolve these issues. 

The current meshless discretization works well for problems in which
the solution does not go negative. Once a negative flux fixup treatment
is applied, the addition of additional physics to the transport discretization
such as thermal radiative transfer and radiation hydrodynamics would
be more achievable. In a radiation hydrodynamics simulation, where
the meshless topology is constantly changing, the consistently discretized
transport with a Voronoi integration approach eliminates mapping to
and from a mesh for radiation transport and is far cheaper than placing
a non partition-of-unity quadrature for each kernel. 

\section*{Acknowledgements}

This work was performed under the auspices of the U.S. Department
of Energy by Lawrence Livermore National Laboratory under Contract
DE-AC52-07NA27344. This document was prepared as an account of work
sponsored by an agency of the United States government. Neither the
United States government nor Lawrence Livermore National Security,
LLC, nor any of their employees makes any warranty, expressed or implied,
or assumes any legal liability or responsibility for the accuracy,
completeness, or usefulness of any information, apparatus, product,
or process disclosed, or represents that its use would not infringe
privately owned rights. Reference herein to any specific commercial
product, process, or service by trade name, trademark, manufacturer,
or otherwise does not necessarily constitute or imply its endorsement,
recommendation, or favoring by the United States government or Lawrence
Livermore National Security, LLC. The views and opinions of authors
expressed herein do not necessarily state or reflect those of the
United States government or Lawrence Livermore National Security,
LLC, and shall not be used for advertising or product endorsement
purposes.

\bibliographystyle{unsrt}
\bibliography{rk_transport}

\appendix

\section{Integral transformations\label{sec:Transformations}}

This section describes transformations of volume and surface integrals
from a reference element (a line segment in 1D, a triangle in 2D,
and a tetrahedron in 3D) to an element in physical space. For information
on the meshless integration methods that produce these elements, see
Sec. \ref{sec:Integration}. 

\subsection{Volume integrals}

The integrals described in Sec. \ref{sec:Integration} need to be
mapped from reference space $R$ to physical space $V$, 
\begin{equation}
\int_{V}f\left(x\right)dV=\int_{R}f\left(x\left(\xi\right)\right)\left|J\right|dR,
\end{equation}
where $x$ and $\xi$ are the coordinates in physical and reference
space, respectively, and $J$ is the Jacobian determinant of the transformation,
\begin{equation}
J^{\alpha\beta}=\frac{\partial x^{\beta}}{\partial\xi^{\alpha}}.
\end{equation}
In discrete form, the quadrature is mapped using 
\begin{equation}
\int_{V}f\left(x\right)dV\approx\sum_{m}w_{m}\left|J_{m}\right|f\left(x\left(\xi_{m}\right)\right),
\end{equation}
where $w_{m}$ are the weights of the reference quadrature, which
effectively converts the ordinates to $x\left(\xi_{m}\right)$ and
the weights to $w_{m}\left|J_{m}\right|$ for the integration. 

The line quadrature is assumed to have the bounds $-1\leq\xi\leq1$,
while the triangular and tetrahedral quadratures have the bounds $0\leq1^{\alpha}\xi^{\alpha}\leq1$,
where $1^{\alpha}$ represents a vector of ones. For a line with points
$p_{0}$ and $p_{1}$ in physical space, the mapping to reference
space is
\begin{equation}
x\left(\xi\right)=p_{0}+\left(p_{1}-p_{0}\right)\frac{\left(\xi+1\right)}{2}.\label{eq:line-map}
\end{equation}
For a triangle with points $p_{0}$, $p_{1}$, and $p_{2}$ or a tetrahedron
with an additional point $p_{3}$, the mapping is 
\begin{equation}
x\left(\xi\right)=p_{0}+\left(p_{\alpha}-p_{0}\right)\xi^{\alpha}.\label{eq:tri-map}
\end{equation}

\subsection{Surface integrals}

For surface integrals, the mapping is between reference space $T$
with coordinates $\xi$ and physical space $S$ with coordinates $x$,
\begin{equation}
\int_{S}f\left(x\right)dS=\int_{T}f\left(x\left(\xi\right)\right)KdT,
\end{equation}
where $K$ is a differential surface element. For a triangle mapped
from a 2D reference element to a 3D surface element, this term is
\begin{equation}
K=\left|\partial_{\xi_{1}}x\times\partial_{\xi_{2}}x\right|,
\end{equation}
while for a line mapped from a 1D to a 2D line element, the differential
element is 
\begin{equation}
K=\left|\partial_{\xi}x\right|,
\end{equation}
with the line and surface integrals mapped as in Eqs. (\ref{eq:line-map})
and (\ref{eq:tri-map}), with the exception that the $p$ vectors
have one more element than the $\xi$ ones (e.g. in 3D, the $p$ vectors
have three elements, while the $\xi$, which is a surface parameterization,
has only two). In discrete form, the surface integrals are 
\begin{equation}
\int_{S}f\left(x\right)dS\approx\sum_{m}w_{m}K_{m}f_{m}.
\end{equation}

\section{Analytic solution to asteroid problem\label{sec:Analytic-asteroid}}

In this section, the analytic solution for the asteroid problem described
in Sec. \ref{subsec:Asteroid} is derived. In 2D, the ``asteroid''
is actually an infinite cylinder and the ``point source'' is a line
source. The transport equation for this problem can be written in
cylindrical geometry as 
\begin{equation}
\mu\frac{1}{r}\partial_{r}\left(r\psi\right)+\sigma_{a}\psi=\delta\left(r\right),
\end{equation}
where $r$ is the distance from the point source and $\mu$ is the
cosine of the angle between the x-y plane and the radiation propagation
direction. The cross section is 
\begin{equation}
\sigma_{a}=\begin{cases}
\sigma_{a,\text{background}}, & r<d_{\text{asteroid}},\\
\sigma_{a,\text{asteroid}}, & \text{otherwise},
\end{cases}
\end{equation}
where $d_{\text{asteroid}}$ is the distance from the source to the
asteroid for a given evaluation point. The solution to this equation
is 
\begin{equation}
\psi=\frac{1}{r}\exp\left(-\frac{1}{\mu}\omega\right),\label{eq:asteroid-angular}
\end{equation}
where 
\begin{equation}
\omega=\int_{0}^{r}\sigma_{a}dr',\label{eq:optical-depth-asteroid}
\end{equation}
which can be calculated using only the distance travelled in each
of the asteroid and the background along with their respective cross
sections. Integrating this equation over all forward angles, $0\leq\mu\leq1$
(which is also the only set of angles for which the solution is nonzero),
the scalar flux solution is 
\begin{equation}
\phi=\frac{1}{r}\left[\exp\left(-\omega\right)-\omega\int_{\omega}^{\infty}\frac{1}{\omega'}\exp\left(-\omega'\right)d\omega'\right].\label{eq:asteroid-scalar}
\end{equation}
The remaining integral is the exponential integral, which can be evaluated
directly in many mathematical software packages. 

To find the distance from the source to the asteroid, let the position
of the source in the $xy$ plane be $x_{\text{src}}$ and the evaluation
point be $x_{\text{eval}}$. The parametric equation for the line
connecting these is 
\begin{equation}
x=x_{\text{src}}+\left(x_{\text{eval}}-x_{\text{src}}\right)s,\label{eq:parametric-s}
\end{equation}
where $s=0$ is located at the source and $s=1$ is at the evaluation
point. Inserting this equation into the equation for the asteroid
centered at the origin, 
\begin{equation}
\left|x\right|^{2}=r_{\text{asteroid}}^{2},
\end{equation}
and solving for $s$ results in the two intercept locations,
\begin{equation}
s_{\text{int}}=\frac{-\ell_{1}\pm\sqrt{\ell_{1}^{2}-\ell_{0}\ell_{2}}}{\ell_{2}},\label{eq:s-intercept}
\end{equation}
where 
\begin{gather}
\ell_{0}=x_{\text{src}}^{\alpha}x_{\text{src}}^{\alpha}-r^{2},\\
\ell_{1}=x_{src}^{\alpha}\left(x_{\text{eval}}^{\alpha}-x_{\text{src}}^{\alpha}\right),\\
\ell_{2}=\left(x_{\text{eval}}^{\alpha}-x_{\text{src}}^{\alpha}\right)\left(x_{\text{eval}}^{\alpha}-x_{\text{src}}^{\alpha}\right).
\end{gather}
If $\ell_{1}^{2}-\ell_{0}\ell_{2}<0$, then the ray from the source
to the evaluation point does not travel through the asteroid. The
evaluation point may be before the ray intersects with the asteroid,
inside the asteroid, or after the ray has exited the asteroid. Given
the distances traveled in each of the asteroid and the background,
the integral in Eq. (\ref{eq:optical-depth-asteroid}) can be evaluated,
which permits evaluation of either the angular or scalar flux {[}Eq.
(\ref{eq:asteroid-angular}) and (\ref{eq:asteroid-scalar}), respectively{]}. 

\pagebreak{}

\begin{algorithm}[H]
\hfill{}%
\begin{minipage}[t]{0.95\columnwidth}%
\begin{lstlisting}[numbers=left,basicstyle={\small\sffamily},breaklines=true,tabsize=2]
class BilinearKernelDKernel : public BilinearIntegral
	func addToIntegral(basis, dbasis, ordinate, weight):
		set numBasis to number of basis functions (neighbors)
		for (i = 0; i < numBasis; ++i)
			for (j = 0; j < numBasis; ++j)
			integral(i, j) += weight * coefficient(ordinate) * basis[i] * basis[j]
	SparseMatrix integral
class BilinearSurfaceKernelKernel : public BilinearIntegral
	func addToSurfaceIntegral(basis, ordinate, weight, normal):
		set numBasis to number of basis functions (neighbors)
		for (i = 0; i < numBasis; ++i)
			for (j = 0; j < numBasis; ++j)
				integral(i, j) += weight * normal.dot(coefficient(ordinate)) * basis[i] * basis[j]
	SparseMatrix integral
func performIntegration(volumeIntegrals, surfaceIntegrals):
	for (i = 0; i < numPoints; ++i) 
		set neighbors to neighbors of point i
		set cell to the voronoi tesselation for the point i
		decompose cell into subcells
		for (c = 0; c < numSubcells; ++c) 
			set ordinates and weights to volume quadrature for subcells[c]
			for (q = 0; q < numOrdinates; ++q) 
				set basis/dbasis to RK evaluations/derivatives at ordinates[q] for each neighbor
				for each volumeIntegral
					integral.addToIntegral(basis, dbasis, ordinates[q], weights[q]) 
		decompose cell surface into subsurfaces
		for (s = 0; s < numSubsurfaces; ++s) 
			set ordinates and weights to surface quadrature for subsurfaces[s]
			set normal to normal for subsurfaces[s]
			for (q = 0; q < numOrdinates; ++q) 
				set basis to RK evaluations for each neighbor point at ordinates[q]
				for each surfaceIntegral:
					integral.addToSurfaceIntegral(basis, ordinates[q], weights[q], normal[s])
\end{lstlisting}
\end{minipage}

\caption{Meshless integration algorithm shown for the example functions $\left\langle \partial_{x}^{\alpha}U_{i},fU_{j}\right\rangle $
and $\left(U_{i},n^{\alpha}g^{\alpha}U_{j}\right)$.}
\label{alg:integration}
\end{algorithm}

\begin{algorithm}[H]
\hfill{}%
\begin{minipage}[t]{0.95\columnwidth}%
\begin{lstlisting}[numbers=left,basicstyle={\small\sffamily},breaklines=true,tabsize=2]
func getRefinedQuadrature(minRule, goalRule, angleHitsObject):
	initialize ordinates and weights to be empty
	get quadrature for the goal rule
	for (rule = minRule; rule < goalRule; ++rule)
		initialize ordinates and weights for this quadrature
		for (i = 0; i < numPoints for this rule; ++i)
			get the indices from the goal quadrature that this point represents as goalPoints
			if not angleHitsObject(goalPoints[j]) for any such goal quadrature index j:
				add this point to the ordinates and weights
				remove this point from the goal quadrature and weights
	add the remaining goal quadrature ordinates and weights to the ordinates and weights
	return the ordinates and weights
\end{lstlisting}
\end{minipage}

\caption{Method for refinement of the angular quadrature to resolve a region
of interest.}
\label{alg:angular-refinement}
\end{algorithm}

\begin{center}
\begin{figure}[H]
\centering{}%
\begin{minipage}[t]{0.47\columnwidth}%
\includegraphics[width=1.0\textwidth,trim={0cm 0cm 0cm 0cm},clip]{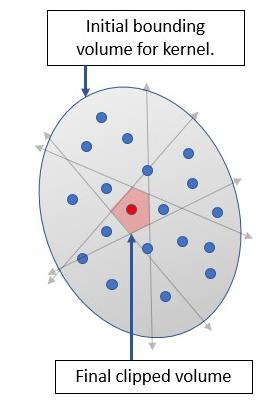}

\caption{To construct the unique volume for the point in red, a polygon is
constructed that bounds the point's non-zero kernel volume. This polygon
is then clipped by the perpendicular half-planes for each point it
interacts with. For overlapping points this results in a polygon equivalent
to the Voronoi tessellation for the point in question. }
\label{fig:polygon-volume}%
\end{minipage}\hfill{}%
\begin{minipage}[t]{0.47\columnwidth}%
\includegraphics[width=1.0\textwidth,trim={0cm 0cm 0cm 0cm},clip]{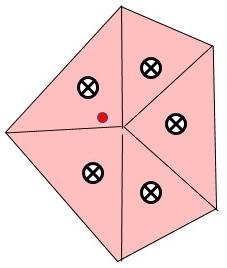}

\caption{Once the polygon for a point is created, it is decomposed into triangles
(2D) or tetrahedra (3D) for integration.}
\label{fig:polygon-decomposition}%
\end{minipage}
\end{figure}
\par\end{center}

\begin{center}
\begin{figure}[H]
\begin{centering}
\includegraphics[width=0.55\textwidth,trim={0cm 0cm 0cm 0cm},clip]{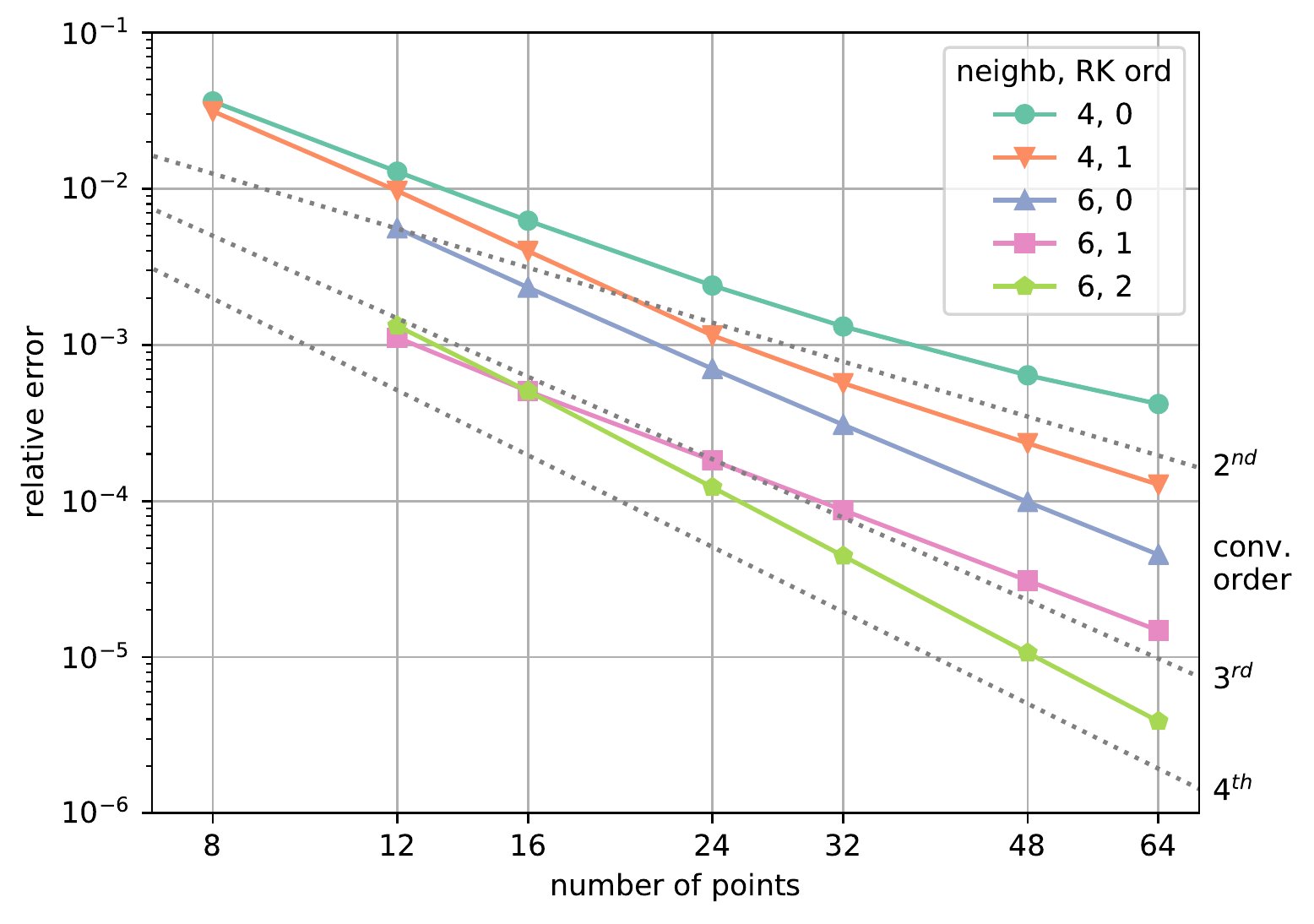}
\par\end{centering}
\caption{Relative error of the numeric solution to the purely absorbing problem
as the number of points is increased for various combinations of the
number of neighbors and RK correction order, with the dotted lines
indicating second, third, and fourth-order convergence.}
\label{fig:purely-absorbing-err}
\end{figure}
\par\end{center}

\begin{center}
\begin{figure}[H]
\begin{centering}
\includegraphics[width=0.55\textwidth,trim={0cm 0cm 0cm 0cm},clip]{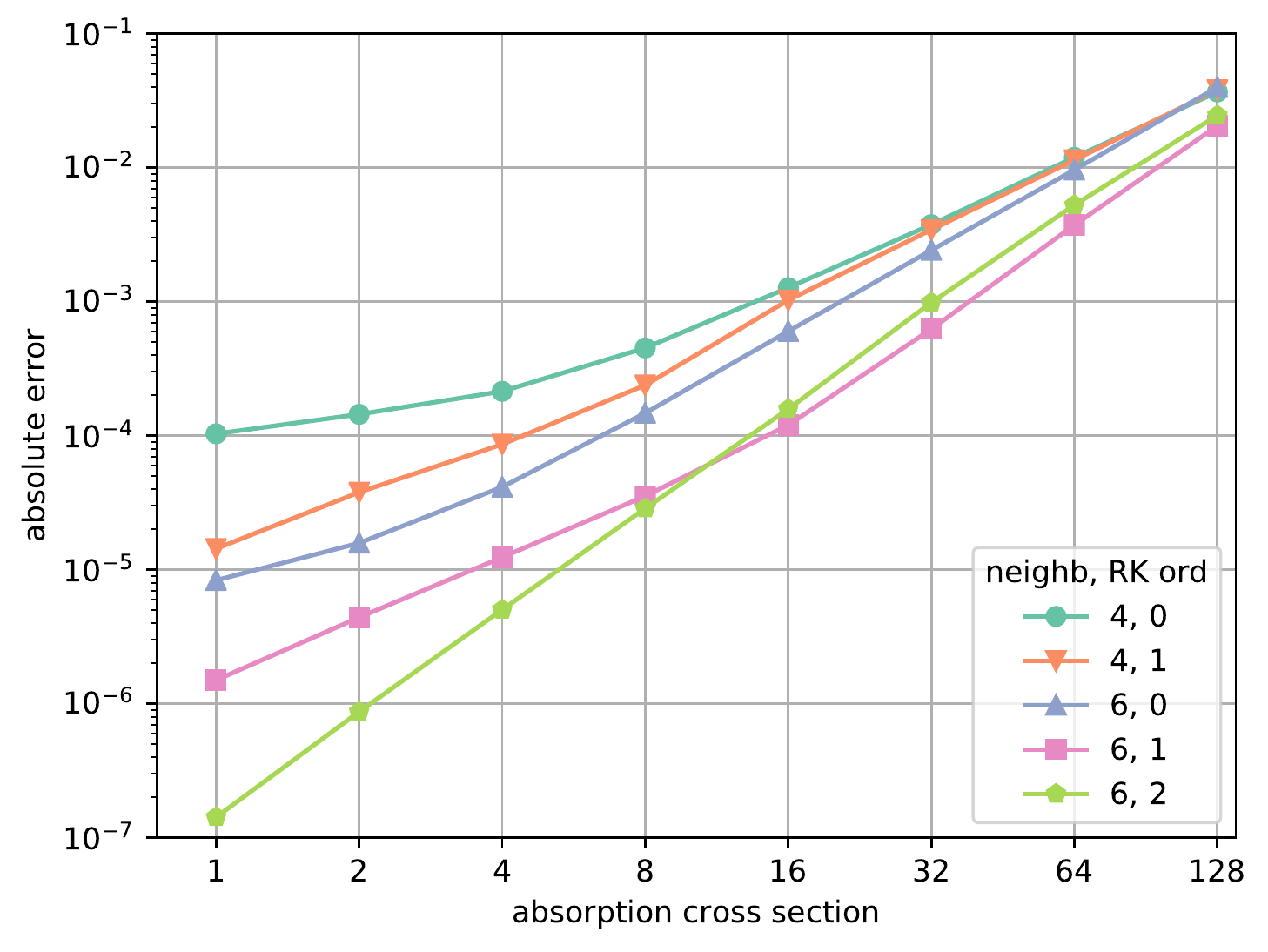}
\par\end{centering}
\caption{Absolute error of the numeric solution to the purely absorbing problem
for various combinations of the number of neighbors, RK correction
order, and absorption cross section.}
\label{fig:purely-absorbing-xs}
\end{figure}
\par\end{center}

\begin{center}
\begin{figure}[H]
\begin{centering}
\includegraphics[width=1.0\textwidth,trim={5cm 4cm 7cm 2cm},clip]{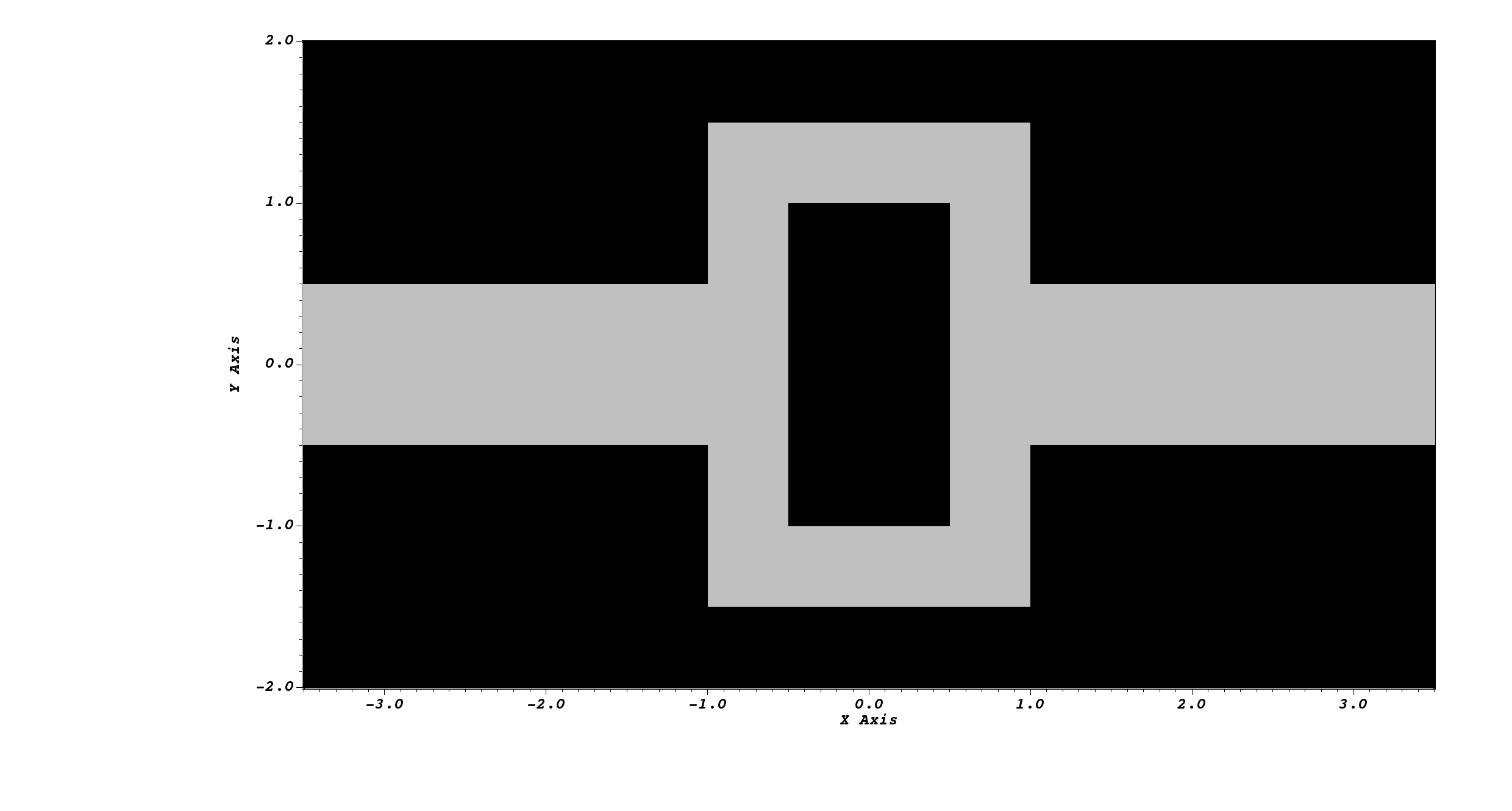}
\par\end{centering}
\caption{Geometry of the cooked pipe problem. The scattering opacities are
$\sigma_{s}=200$ in the black region and $\sigma_{s}=0.2$ in the
gray region.}
\label{fig:crooked-geometry}
\end{figure}
\par\end{center}

\begin{center}
\begin{figure}[H]
\begin{centering}
\includegraphics[width=1.0\textwidth,trim={5cm 4cm 7cm 2cm},clip]{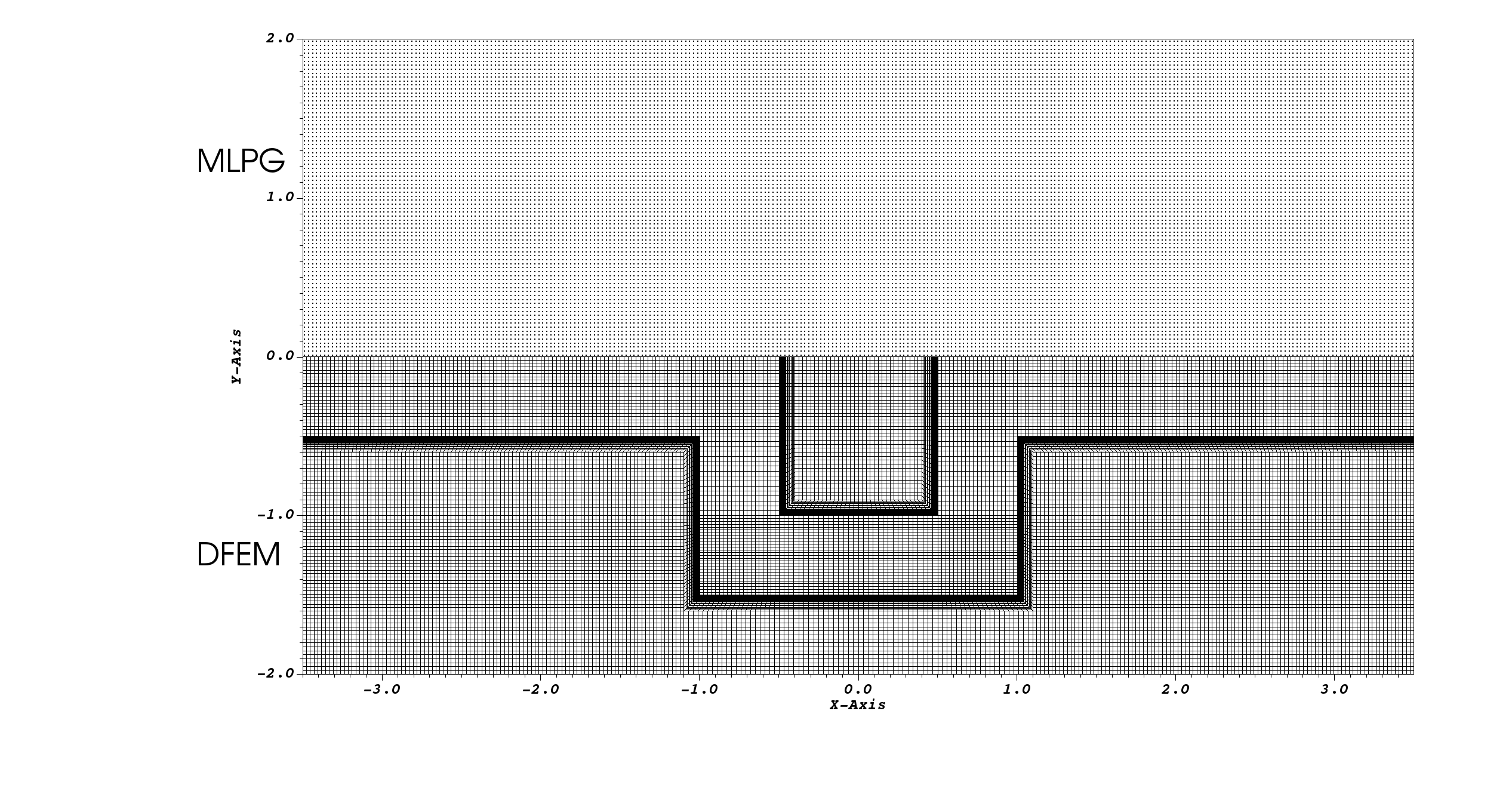}
\par\end{centering}
\caption{The MLPG points (top) and DFEM mesh (bottom) for the crooked pipe
problem, at 1/4 resolution for visibility. Note that the DFEM mesh
has additional resolution near the thin-thick boundary.}
\label{fig:crooked-mesh}
\end{figure}
\par\end{center}

\end{document}